\documentclass[journal]{IEEEtran}

\usepackage{moreverb,url}

\usepackage{graphicx,epstopdf,amsthm,amssymb,url,amsmath,color,mathtools,epsfig,float} %
\usepackage{algorithm,algorithmic,graphicx,times,amsbsy,bm,amsmath,caption}

\usepackage{float,paralist,epstopdf,amsmath,empheq}
\def\mathbi#1{\textbf{\em #1}}
\graphicspath{ {./Figures/}}
\DeclareMathAlphabet\mathbfcal{OMS}{cmsy}{b}{n}
\usepackage[keeplastbox]{flushend}

\usepackage{cite}

\begin{document}

\title{Matrix-Based Characterization of the Motion and Wrench Uncertainties in Robotic Manipulators}

\author{Javad~Sovizi,
	Sonjoy~Das
	and~Venkat~Krovi
\thanks{J. Sovizi was with the Department
		of Mechanical and Aerospace Engineering, University at Buffalo, Buffalo,
		NY, 14260 USA e-mail: javadsov@buffalo.edu.}
\thanks{S. Das is with the Department
	of Mechanical and Aerospace Engineering, University at Buffalo, Buffalo,
	NY, 14260 USA.}
\thanks{V. Krovi is with the Departments of Automotive Engineering and Mechanical Engineering, Clemson University, Greenville, SC 29607 USA.}}

\maketitle 
%
%

\begin{abstract}
Characterization of the uncertainty in robotic manipulators is the focus of this paper. Based on the random matrix theory (RMT), we propose uncertainty characterization schemes in which the uncertainty is modeled at the macro (system) level. This is different from the traditional approaches that model the uncertainty in the parametric space of micro (state) level. We show that perturbing the system matrices rather than the state of the system provides unique advantages especially for robotic manipulators. First, it requires only limited statistical information that becomes effective when dealing with complex systems where detailed information on their variability is not available. Second, the RMT-based models are aware of the system state ({\it e.g.}, joint speeds) and configuration ({\it e.g.}, close-to-singularity) that are significant factors affecting the level of uncertainty in system behavior. 
In this study, in addition to the motion uncertainty analysis that was first proposed in our earlier work, we also develop an RMT-based model for the quantification of the static wrench uncertainty in multi-agent cooperative systems. This model is aimed to be an alternative to the elaborate parametric formulation when only rough bounds are available on the system parameters. We discuss that how RMT-based model becomes advantageous when the complexity of the system increases. We perform experimental studies on a robotic manipulator (5DOF KUKA youBot arm) to demonstrate the superiority of the RMT-based motion uncertainty models. We show that how these models outperform the traditional models built upon Gaussianity assumption in capturing real-system uncertainty and providing accurate bounds on the state estimation errors. In addition, to experimentally support our wrench uncertainty quantification model, we study the behavior of a cooperative system of the mobile robots (multiple iRobots). It is shown that one can rely on less demanding RMT-based formulation and yet meets the acceptable accuracy.  	      
\end{abstract}



\section{Introduction}
\label{intro}
Application of robotic manipulators is continuously advancing toward increasingly complex tasks that require high levels of reliability and safety. Some representative examples of such applications are minimally invasive surgeries, micro- and nano-manipulations and cooperative payload transportation, among many others. The complexity of the task often leads to the complexity in kinematic structure of the robotic system. As a result, the modeling uncertainty (due to all ignored/unknown factors that contribute to deviation of the actual response from deterministic predictions) increases leading to a poor fidelity of the stochastic models of the system. In order to properly characterize (capture) the variation of a particular system, it is necessary to have: (i) a descriptive mathematical model of the physical system that embraces all the effective variables; (ii) statistical information on the random variation of all the variables; and (iii) a reliable and efficient technique to propagate the uncertainty. While the last element can be addressed by one of the many uncertainty quantification techniques that pass the uncertainty through the nonlinear model of the system, the first two elements of the characterization platform are often difficult or intractable to achieve. There is a plethora of the methods reported in the literature where detailed formulation of the governing equations is proposed in order to capture the effect of different factors ignored in the simpler models. However, there are always unidentified factors that may affect the system response. In addition, incorporating all the effects that are known to the designer typically results in a highly nonlinear and complex model to which the solution is computationally demanding. Moreover, characterization (finding the statistical information) of each single variable requires several experimental analyses that can not be pursued due to the cost and time needed for such detailed investigations. To this end, in this paper we propose and examine system-level approaches based on the random matrix theory (RMT) that provide several advantages over the parametric uncertainty characterization models. 

The idea behind the system-level RMT-based models is in fact the consistent modeling of the uncertainty  at the macro-level rather than modeling the system state level (micro-level) parameters. This requires only limited information for formulating a probabilistic model of the system. In addition, system level formulations are aware of the system state and configuration that play important role in response variation. For example, formulating the Jacobian matrix as a random matrix, that is the focus of this paper, provides information to the probabilistic models of the state and configuration (such as close-to-singularity configurations). Finally, macro-level formulations provide computational efficiency in Monte Carlo based approaches by directly sampling the system matrices and eliminating the need for propagating the uncertainty jointly induced by several random variables at the micro-level.     

Uncertainty modeling in robotic systems dates as far back as work by Smith and Cheeseman \cite{Smith} in 1986. They developed a first order model for the propagation of the uncertainty under a series of the frame transformations, given the statistics of the uncertainty associated with a single transformation. Following their work, several researchers performed different studies that are mainly focused on the simultaneous localization and map building (SLAM) \cite{Dissan, Mont, Simmons, DURRANT} in the mobile robotic systems. The problem of uncertainty propagation focusing particularly on manipulator systems has been elegantly addressed by Wang and Chirikjian \cite{WangChirik2,WangChir1}. The error propagation problem on the Euclidean motion groups is considered in their work where the probability density function (pdf) of the end-effector pose is derived using the convolution of the densities corresponding to each unit along the serial chain. They also extended  their method \cite{wang2008nonparametric} to the second order approximation of larger error propagation using the theory of Lie algebras and Lie groups. 

Despite several works addressing the uncertainty in robotic systems, the literature is mainly biased toward quantification of the uncertainty propagation. In this paper, we first review our RMT-based formulation of the robotic system matrices (\textit{e.g.}, Jacobian and mass matrices) proposed in our earlier studies \cite{sovizi2013random, SoviziICRA, sovizi2013uncertainty, SoviziASME, sovizi2016application, sovizi2014random}. Two models are proposed for formulating the Jacobian matrix: first model is mainly adopted from the structural analysis literature developed by Soize \cite{Soize0, Soize1, Soize2} and second model is developed based on Gaussianity assumptions and a measure of the manipulator dexterity. We believe and experimentally show that the proposed RMT-based models can be integrated into uncertainty quantification techniques to provide further improved frameworks to address the uncertainty in the design and control of the robotic systems.  

Moreover, we propose an RMT-based wrench uncertainty model to formulate the uncertainty in complex multi-agent systems that consist of several simple robotic agents.  We are aiming to provide a probabilistic formulation that properly captures the wrench uncertainty despite the complexity of the system, and enables the integration of uncertainty level into general motion planning and optimal control strategies. Addressing this problem becomes important in a variety of complex systems such as rehabilitation \cite{alamdari2016design, ghannadi2015development} and surgical robotic systems \cite{wilkening2017, alambeigi2017, alambeigi2016toward} where the information on the statistics of output wrench becomes crucial for the safety and achieving the desired outcome. However, in this work we are particularly interested in analyzing cable-based multi-agent cooperative systems. The main idea behind such systems design is to distribute a task, unachievable for a single agent, among several agents that may or may not be identical. While several advantages are provided by such flexible and fault-tolerant systems, redundancy introduces many challenges, one of which is systematically allocating the tasks to different agents in order to optimize some performance indices. Such problems have received significant attention from researchers \cite{berman2009,fink2008,cheng2009,matthey2009} especially in the context of decentralized control and optimal path planning.

\begin{figure}[htb!]
	\centering
	\begin{minipage}[t]{0.25\textwidth}
		\includegraphics[trim=70 50 50 70, clip, scale=0.16]{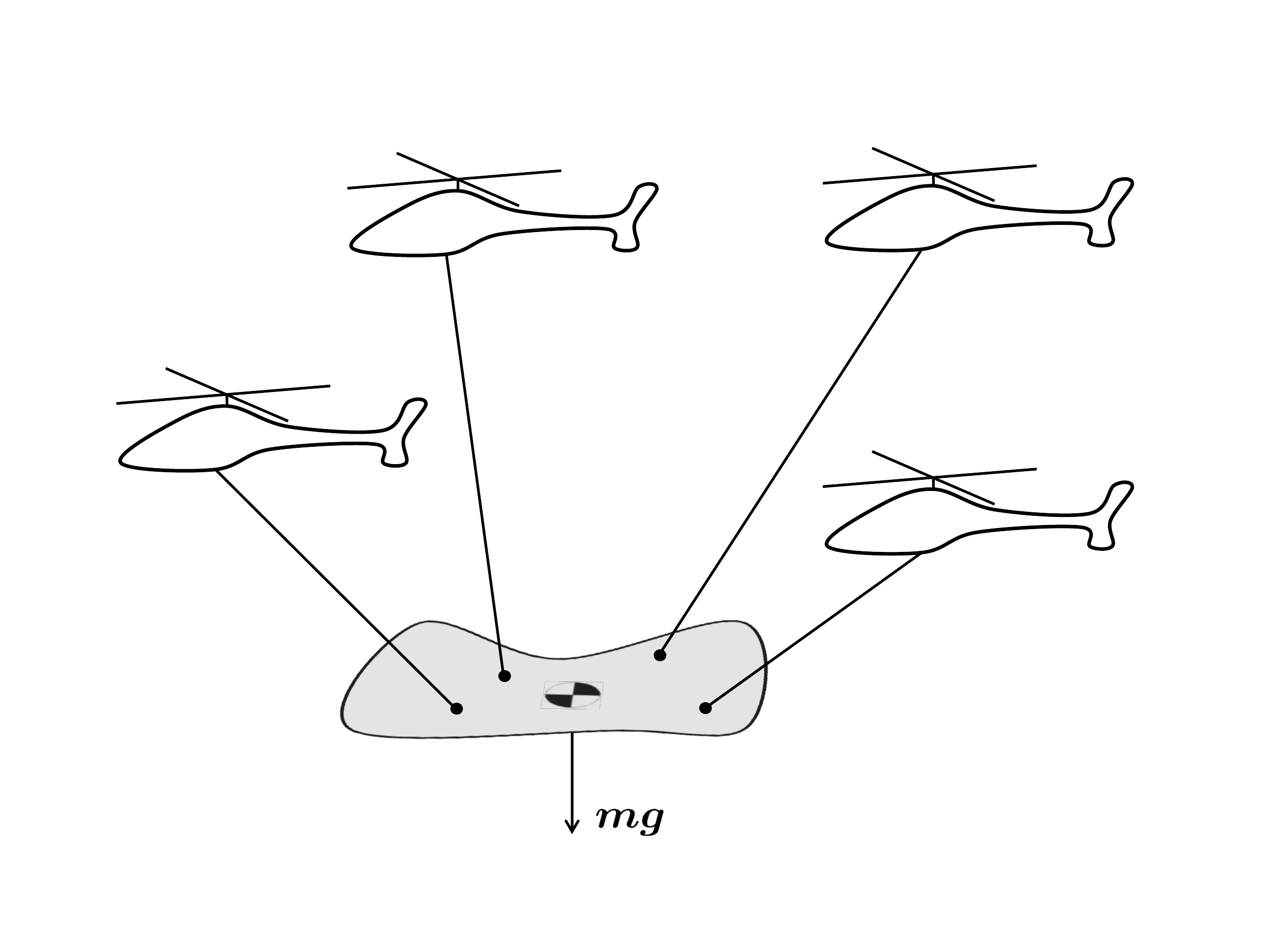}
		
		\hspace{0.55in}(a)
	\end{minipage}%
	\hspace{-0.4in}
	\begin{minipage}[t]{0.25\textwidth}
		\includegraphics[trim=50 50 50 70, clip, scale=0.176]{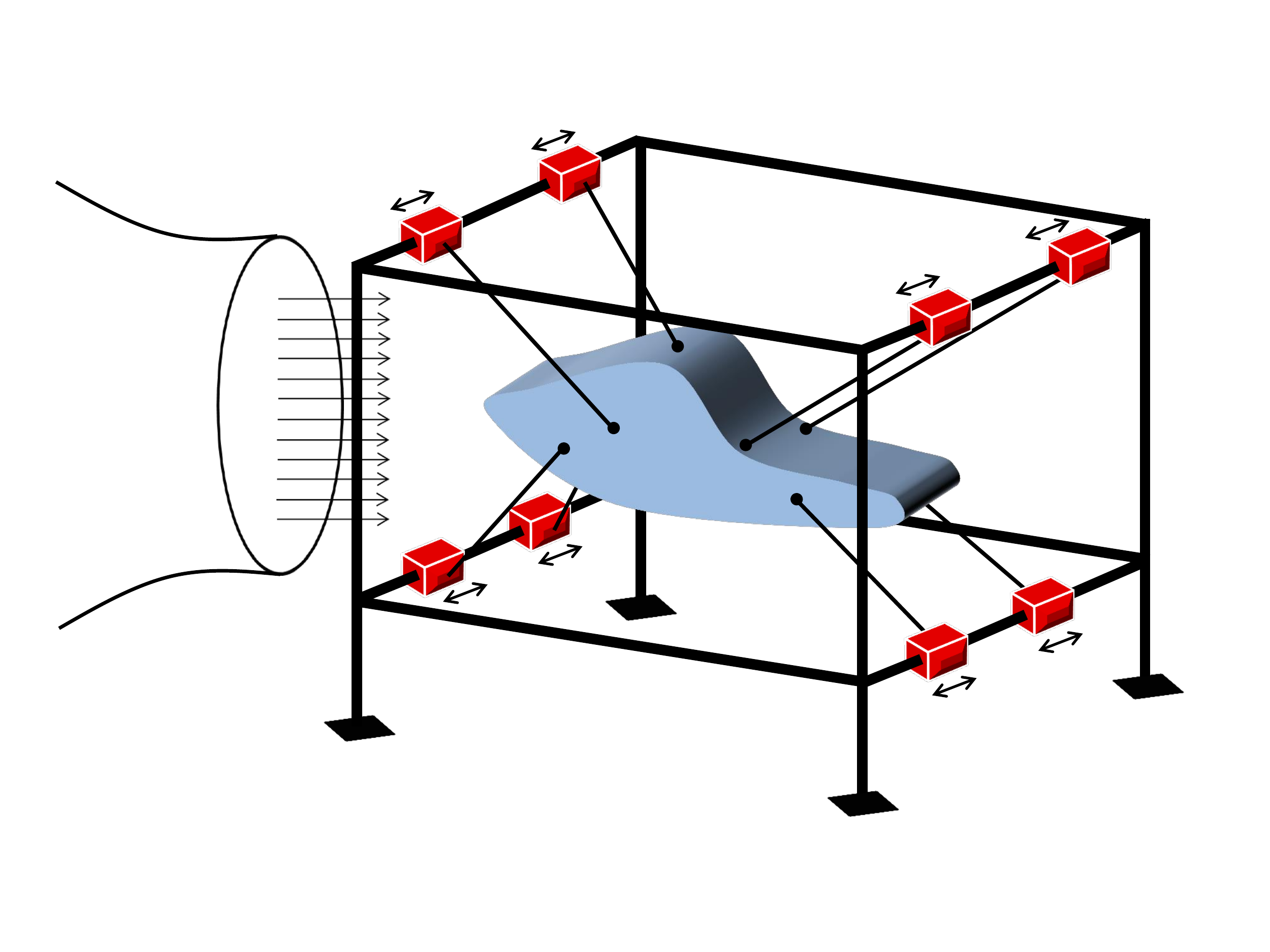} 
		
		\hspace{0.65in}(b)
	\end{minipage}%
	\caption{Representative examples of the cooperative systems: (a) aerial towing/suspension using flying robots; (b) airplane hangar where bases can be re-adjusted along the rails.}
	\label{Exmpls}\vspace{-0.2in}
\end{figure}

Special lay out of such cooperative systems makes them vulnerable to (i) configuration/calibration uncertainty and (ii) actuation/sensing uncertainty that ultimately result in an uncertain output wrench. Two representative examples of these systems are shown in Fig.~\ref{Exmpls}.
 For instance, in aerial towing problem (Fig.~\ref{Exmpls} (a)),  flying robots (such as commonly used quadcopters) are inherently uncertain in motion (especially when loaded) that leads to a highly uncertain configuration. A comprehensive study of such systems can be found in the literature \cite{FinkKumarIJRR2010,MichaelKumarAutRob2011,Geom2013Kumar,DynPlanKumar2013,InvKin2013Kumar,goodarzi2015}. Uncertain position of the flying robots has been reported to be responsible for control errors (for example, see \cite{MichaelKumarAutRob2011}). Similarly in measurement systems such as aircraft hangar \cite{Lafou2002,Yang2010} (shown in Fig.~\ref{Exmpls} (b)), configuration of the agents is uncertain due to the calibration errors, flexibility and displacement of the bases, etc. Furthermore, actuation and sensing uncertainties impact the performance of the systems shown in Fig.~\ref{Exmpls} (a) and (b), respectively. In a wrench-delivery (forward) problem such as aerial towing, there exist an intrinsic imperfection of the actuators in providing the exact (commanded) force. Similarly in a wrench-sensing (inverse) problem such as platform wrench measurement shown in Fig.~\ref{Exmpls} (b), uncertainty in sensing the force/tension corresponding to each individual agent (cable tension) is inevitable. 
 
We incorporate these uncertainties in a static formulation of the output wrench. In a wide range of applications, the operation condition is such that a quasi-static analysis can significantly characterize the system response (for example, see \cite{DynPlanKumar2013}). Hence, our formulation is applicable to several physical systems working in such conditions. In an RMT-based framework and based on some Gaussianity assumptions, we construct a model and provide the closed-form expression of the resultant wrench uncertainty level (output wrench covariance). Such closed-form expressions become very useful in design and real-time motion planning for such systems when the optimization algorithm searches for the configurations that minimize the uncertainty in system response. Conventional cable-robots use fixed bases with varying cable lengths \cite{Albus1993, AlpAgrawal2002, FattAgraw2005, alamdari2015robotic, alamdari2015parallel, alamdari2016design}, however, reconfigurable systems where the cable lay out can be changed through mobility of the bases have been developed in recent studies \cite{Merlet2010, Rosati2011, XiaoboICRA12, Xiaobo2013, NguyenICRA14, NguyenIROS14, anson2017orientation}. Reconfiguration planning in such systems can benefit from formulations proposed in this paper by augmenting the objective function with covariance terms that can be computed efficiently. It also facilitates analytical investigation and systematic trade-offs in multi-agent systems design to study the sensitivity of the response uncertainty to different design parameters. The RMT-based model, proposed in this paper, is an alternative to the detailed parametric formulation that we provided in our recent study \cite{SoviziTRO}. We show that elaborate and (roughly speaking) information-demanding parametric models can be replaced by less demanding RMT-based models at the expense of reducing the accuracy. Under many circumstances and especially in multi-agent systems with large number of cooperative robots, the loss of accuracy due to utilizing RMT-based models can be sufficiently low such that the fidelity of the model is yet acceptable.

The rest of this paper is organized as follows. Next section presents some preliminaries required for RMT-based formulations. In Sec.~\ref{RMTFor}, RMT-based uncertainty formulations are developed. First we present the approaches for motion uncertainty analysis: maximum entropy (MaxEnt) approach and Gaussian model. Next in subsection~\ref{GaussProd}, we develop the RMT-based model for wrench uncertainty quantification in multi-agent systems. Section~\ref{ExpRes} includes our experimental analyses for both the motion uncertainty (KUKA youBot arm) and output wrench  uncertainty (multiple iRobots) analyses. Finally, discussion and the direction of future work are provided in Sec.~\ref{Discuss}.  
\section{Preliminaries}
\label{Prelmny}
Relying on the existing literature \cite{Gupta}, we present the theorems and definitions that are used throughout the paper to develop the proposed RMT-based uncertainty characterization models.  
 
\textbf{Definition 1}: Random matrix $\bm{X}$ ($\bm{X}\in\mathbb{M}_{p,q}$ where $\mathbb{M}_{p,q}$ is the set of all real $p \times q$ matrices) is said to have a matrix-variate Gaussian distribution with mean matrix $M(p\times q)$ and covariance matrix $\Sigma\otimes\Psi$ where $\Sigma(p\times p)>0$ ($A>0$ if $A$ is positive definite matrix) and $\Psi(q\times q)>0$, denoted as $\bm{X}\sim N_{p,q}(M,\Sigma\otimes\Psi)$, if $\mbox{vec}(\bm{X}^{T})$ has multi-variate normal distribution with mean $\mbox{vec}(M^{T})$ and covariance matrix $\Sigma\otimes\Psi$, denoted as $\mbox{vec}(\bm{X}^{T})\sim N_{pq}(\mbox{vec}(M^{T}),\Sigma\otimes\Psi)$. 

In Definition 1, 
\begin{equation}
\mbox{vec}(A_{p\times q})=\begin{bmatrix}
a_{1}\\
\vdots\\
a_{q}
\end{bmatrix}
\end{equation}
where $a_{i}, \ i=1, \dots, q$ is the $i^{th}$ column of $A$. Moreover, $\otimes$ represents Kronecker products defined by
\begin{equation}
A\otimes B=\begin{bmatrix}
a_{11}B & a_{12}B & \dots & a_{1n}B\\
a_{21}B & a_{22}B & \dots & a_{2n}B\\
\vdots & & &\\
a_{m1}B & a_{m2}B & \dots & a_{mn}B
\end{bmatrix}
\label{Kron}
\end{equation}

The density function of $\bm{X}$ is given by
\begin{align}
p_{\bm{X}}(X) &= (2\pi)^{-\frac{1}{2}qp}|\Sigma|^{-\frac{1}{2}q}|\Psi|^{-\frac{1}{2}p}\nonumber\\
& \mbox{etr}\Big\{-\frac{1}{2}\Sigma^{-1}(X-M)\Psi^{-1}(X-M)^{T}\Big\}
\label{GaussMatr}
\end{align}  
where $\mbox{etr}(.)=\exp\{\mbox{tr}(.)\}$.

\textbf{Definition 2}: Random matrix $\bm{S}$ ($\bm{S} \in \ \mathbb{M}^{+}_{p}$ or $\bm{S}>0$ where $\mathbb{M}^{+}_{p}$ is the set of all $p \times p$ symmetric positive definite matrices), is said to have a Wishart distribution with parameters $d$ and $\Sigma \in \ \mathbb{M}^{+}_{p}$, denoted as $S\sim W_{p}(d,\Sigma)$, if its probability density function is given by 
\begin{align}
& p_{S}(S)=\bigg\{2^{\frac{1}{2}dp}\Gamma_{p}\left(\frac{1}{2}d\right)|\Sigma|^{\frac{1}{2}d})\bigg\}^{-1}\nonumber\\
&|S|^{\frac{1}{2}(d-p-1)}\mbox{etr}\left(-\frac{1}{2}\Sigma^{-1}S\right) 
\label{Wishart}
\end{align}
\noindent
where $S\in \ \mathbb{M}^{+}_{p}$ and $d\geq p$.

\textbf{Theorem 1}: Let $\bm{X}\sim N_{p,q}(M,\Sigma\otimes\Psi)$, then $\bm{X}^{T}\sim N_{q,p}(M^{T},\Psi\otimes\Sigma)$.

\textbf{Theorem 2}: Let $\bm{X}\sim N_{p,q}(\bm{0},\Sigma\otimes I_{q})$, $q\ge p$, then $XX^{T}>0$ with probability one.

\textbf{Theorem 3}: Let $\bm{X}\sim N_{p,q}(\bm{0},\Sigma\otimes I_{q})$ and define $S=XX^{T}$, $q\ge p$, then $S\sim W_{p}(q,\Sigma)$.

\textbf{Theorem 4}: Let $\bm{S}\sim W_{p}(d,\Sigma)$, then $E\left[\bm{S}\right]=d\Sigma$.

\textbf{Theorem 5}: Let $\bm{X}\sim N_{p,q}(M,\Sigma\otimes\Psi)$ and $A$ be a $q\times q$ constant matrix, then $E\big[\bm{X}A\bm{X}^{T}\big]=\mbox{tr}\big\{A^{T}\Psi\big\}\Sigma+MAM^{T}$.
\section{Random matrix based uncertainty characterization} 
\label{RMTFor}
\subsection{Motion uncertainty analysis}
\label{KinAnaly}
Here, we present the RMT-based formulation of the manipulator Jacobian matrix developed in our earlier studies \cite{SoviziASME, SoviziICRA}. In a system with motion uncertainty, the inverse differential kinematic equation can be considered as a general stochastic differential equation given by
\begin{equation}
\frac{d\bm{q}}{dt}=f(\bm{q},\bm{\omega},t)
\label{stochdiffeq}
\end{equation}
\noindent
where $\bm{q}$ is the vector of random joint states, $t$ is the time, $f$ is a vector-function describing the kinematic structure of the manipulator and $\bm{\omega}$ is the process noise. In the discrete time domain, Eq.~(\ref{stochdiffeq}) can be treated as a stochastic difference equation. In the special case of the additive independent noise, one can write 
	\begin{equation}
  \bm{q}_{k+1}=f(\bm{q}_{k})+\bm{\omega}_{k+1}=\bm{q}_{k}+J(\bm{q}_{k})^{-1}\dot{x}^{d}_{k}\Delta t+\bm{\omega}_{k+1}
  \label{InvDiffRndm}
  \end{equation}
where $J(.)$ is the manipulator Jacobian matrix, $\dot{x}^{d}_{k}$ is the desired end-effector velocity at time $t_{k}$ and $\Delta t=t_{k+1}-t_{k}$ is the constant time step. However, due to the facts discussed in Sec.~\ref{intro}, we consider the stochastic difference equation as 
\begin{equation}
  \bm{q}_{k+1}=\bm{q}_{k}+\bm{J}^{-1}_{k}\dot{x}^{d}_{k}\Delta t
  \label{RMTInvDiff}
\end{equation}
where $\bm{J}_{k}$ is the random Jacobian matrix by which the uncertainty is introduced to the system at time $t_{k+1}$. In fact, uncertainty of the state in subsequent time instant, {\it i.e.}, $\bm{q}_{k+1}$ is due to (i) propagation of the uncertainty of $\bm{q}_{k}$, and (ii) the uncertainty introduced by the process noise at time $t_{k+1}$. In contrast with the model in Eq.~(\ref{InvDiffRndm}) that perturbs the joint states by an additive noise, we perturb the kinematic structure of the system by formulating the manipulator Jacobian as random matrix. This ultimately adds a perturbation to the states at $t_{k+1}$, however, it also includes system level effects when we consistently perturb the Jacobian rather than the joint states. For instance, in low-dexterity configurations ({\it e.g.}, close-to-singularity), perturbation of the Jacobian results in introducing higher level of uncertainty compared with high-dexterity regions where manipulator provides improved precision. Moreover, end-effector desired velocity (and hence the required joint velocities) is a significant factor that affects the uncertainty level and is taken into account by perturbing the Jacobian matrix. On top of such advantages that typically result in improved fidelity of the model, the fact that RMT-based model is less-information demanding motivates development and application of such models to several complex systems. We discuss the benefits of the model given by Eq.~(\ref{RMTInvDiff}) in further details in experimental validation section (Sec.~\ref{KUKAKinEx}).  
 
In this section, we are interested in characterizing $p(\bm{J}_{k}|\bm{q}_{k}=q_{k})$ that facilitates statistical formulation of the subsequent configuration of the robot (joint states) when its current configuration is accurately observed. 
\subsubsection{Maximum entropy formulation}
\label{maxEnt}
In order to construct $p(\bm{J}_{k}|\bm{q}_{k}=q_{k})$ using maximum entropy principle, we adopt the approach developed by Soize \cite{Soize0,Soize1} to construct the random symmetric positive definite matrices of the structural systems. The $n \times n$ random symmetric and positive definite matrix $\mathbi{A}$ ($\mathbi{A} \in \ \mathbb{M}^{+}_{n}$) is written as
\begin{equation}
\mathbi{A}={L}^{T}_{\underline{A}}\mathbi{G}_{\mathbi{A}}{L}_{\underline{A}}
\label{SoizeDecom}
\end{equation}
\noindent
where ${L}_{\underline{A}}$ is an upper triangular matrix corresponding to Cholesky decomposition of $\underline{A}$ which is the mean of random matrix $\mathbi{A}$, and $\mathbi{G}_{A}$ is a random symmetric positive definite matrix ($\mathbi{G}_{\mathbi{A}} \in \ \mathbb{M}^{+}_{n}$) with identity mean.
In case of manipulator Jacobian matrix, we replace Cholesky decomposition with LU or QR decompositions (depending on the redundancy in the system). Random Jacobian matrix is written as 
\begin{equation}
\bm{J}_{k}={\underline{J}_{1}}_{k}\bm{B}{\underline{J}_{2}}_{k}
\label{LUDecom}
\end{equation}
\noindent
where ${\underline{J}_{1}}_{k}\in \mathbb{M}_{n}$ and ${\underline{J}_{2}}_{k}\in \mathbb{M}_{n}$ ($\mathbb{M}_{n,n}$ is simply denoted by $\mathbb{M}_{n}$) are obtained through decomposition of the mean Jacobian matrix at $t_{k}$, {\it i.e.}, $\underline{J}_{k}={\underline{J}_{1}}_{k}{\underline{J}_{2}}_{k}$. We use LU decomposition in our simulation in experimental validation section. One may use alternative decomposition approaches such as QR decomposition when dealing with redundant systems with generally rectangular Jacobian matrix. Moreover, in Eq.~(\ref{LUDecom}) $\bm{B}\in \ \mathbb{M}^{+}_{n}$ and $\bm{J}_{k}\in \ \mathbb{M}_{n}$. We assume that the task is defined such that the Jacobian matrix remains non-singular during the robot motion, however, manipulator can work in close-to-singularity regions where the dexterity reduces.  

Maximum entropy (MaxEnt) approach can be used to construct the pdf of $\bm{B}$. The MaxEnt formulation is  
\begin{align}
& \mbox{Maximize} \ \ S(p)=-\int_{\mathbi{B}\in \mathbb{M}^{+}_{n}}p_{\mathbi{B}}(B)\ln\{p_{\mathbi{B}}(B)\}\mbox{d}\mathbi{B} \label{Maxenmeas}\\
& \mbox{s.t.} \nonumber \\
& \hspace{0.25in} \int_{\mathbi{B}\in \mathbb{M}^{+}_{n}}p_{\mathbi{B}}(B)\mbox{d}\mathbi{B}=1 \label{normcnst}\\
& \hspace{0.25in} E\left[\mathbi{B}\right]=\int_{\mathbi{B}\in \mathbb{M}^{+}_{n}}\mathbi{B}p_{\mathbi{B}}(B)\mbox{d}\mathbi{B}=\underline{B}=I_{n}\label{meancnst}\\
& \hspace{0.25in} E[\| \mathbi{B}^{-1} \|_{\mathrm{F}}^{\gamma}] < \infty
\label{invmomnt}
\end{align}
\noindent 
where $S(p)$ is the entropy of $p_{\mathbi{B}}(.)$. Eq.~(\ref{normcnst})-(\ref{invmomnt}) are the normalizing, mean and inverse moment constraints, respectively. Every proper density function satisfies normalizing constraint, {\it i.e.}, integrates to one. The mean constraint in Eq.~(\ref{meancnst}) is from prior knowledge of the nominal system. We have $E[\bm{J}_{k}]=\underline{J}_{k}={\underline{J}_{1}}_{k}{\underline{J}_{2}}_{k}$, and from Eq.~(\ref{LUDecom}), $E[\mathbi{J}_{k}]=E[{\underline{J}_{1}}_{k}\bm{B}{\underline{J}_{2}}_{k}]={\underline{J}_{1}}_{k}E[\bm{B}]{\underline{J}_{2}}_{k}
$ that implies $E[\bm{B}]=I_{n}$. Eq.~(\ref{invmomnt}) describes the inverse moment constraint \cite{Soize0,Soize1} that controls the level of uncertainty by choosing $\gamma$, order of the inverse moment. It can be shown \cite{Soize0,Soize1,Adhikari, AdhikWish} that the solution to the problem described by Eqs.~(\ref{Maxenmeas})-(\ref{invmomnt}) is a pdf from the Wishart distribution family with parameters $d=\theta+n+1$ and $\Sigma=\frac{\displaystyle \underline{B}}{\displaystyle \theta+n+1}$ where $\theta=2\gamma$. Further, in order to properly choose $\theta$ (uncertainty level) Soize \cite{Soize0} introduced the normalized standard deviation as 
\begin{equation}
\sigma_{\mathbi{B}}^{2}=\frac{\displaystyle E[\| \mathbi{B}-E[\mathbi{B}]\|_{\mathrm{F}}^{2}]}{\displaystyle \|E[\mathbi{B}]\|_{\mathrm{F}}^{2}}
\label{NormStd}
\end{equation}
\noindent
where $\sigma_{\mathbi{B}}$ is the dispersion parameter of the random matrix $\mathbi{B}$. Once $\sigma_{\mathbi{B}}$ is chosen using prior knowledge or from experiments (discussed in further detail in Sec.~\ref{ExpRes}), $\theta$ can be obtained by \cite{Adhikari}  
\begin{equation}
\theta=\frac{\displaystyle 1}{\displaystyle \sigma_{\mathbi{B}}^{2}}\left(1+\frac{\displaystyle \{\mathrm{tr}\left({\underline{B}}\right)\}^{2}}{\displaystyle \mathrm{tr}\left({\underline{B}^{2}}\right)}\right)-(n+1)
\label{Thetaa}
\end{equation}
Refer to the literature \cite{Soize0,Soize1,Adhikari, AdhikWish} and \cite{boundedDas,DasThes,DasHybr} for further details on the MaxEnt formulation of the system matrices. The procedure to generate the samples of the process $\bm{q}$ based on the model described in this section is summarized in Algorithm~\ref{WishMCAlg}.  

\begin{algorithm}
    \caption{Monte Carlo simulation of $\bm{J}_{k}$. \textbf{Inputs}: $q_{k}$, $\sigma_{\bm{B}}$
    and $n$; \textbf{Output}: Sample of $\bm{J}_{k}$ (${J}_{k}$) and sample of $\bm{q}_{k+1}$ (${q}_{k+1}$)}\label{WishMCAlg}
    \begin{algorithmic}[1]
		  \STATE Substitute $q_{k}$ in the $J(q_{k})$ to obtain $\underline{J}_{k}$
			\STATE Calculate $\theta$ from Eq.~(\ref{Thetaa})
			\STATE Set $d=\theta+n+1$ and $\Sigma=\frac{I_{n\times n}}{\theta+n+1}$
			\STATE Generate a sample ${B}$ of $\bm{B}$ (MATLAB's command \verb+wishrnd+ may be useful here)
			\STATE Calculate ${\underline{J}_{1}}_{k}$ and ${\underline{J}_{2}}_{k}$ by decomposing $\underline{J}_{k}$
			\STATE Substitute ${\underline{J}_{1}}_{k}$, ${B}$ and ${\underline{J}_{2}}_{k}$ in Eq.~(\ref{LUDecom}) to obtain a sample ${J}_{k}$ of $\bm{J}_{k}$
			\STATE Substitute ${J}_{k}$ in Eq.~(\ref{RMTInvDiff}) to obtain a sample of $\bm{q}_{k+1}$
    \end{algorithmic}
  \end{algorithm}
\subsubsection{Gaussian Jacobian matrix formulation}
MaxEnt formulation of the Jacobian matrix can be effective in characterizing the motion uncertainty (we will discuss this shortly in the Sec.~\ref{KUKAKinEx}). However, it requires a proper decomposition and ultimately provides the pdf of the perturbation matrix rather than Jacobian itself. More specifically, the perturbation matrix represents the underlying fluctuation/perturbation that equivalently perturbs any system regardless of its kinematic structure and once used in the model described by Eq.~(\ref{RMTInvDiff}), the system-specific uncertainty is resulted. While the distribution of the perturbation matrix facilitates sampling $\bm{q}$, further analytical manipulations to find the closed-form expressions that quantify $\bm{q}_{k}$ uncertainty level are difficult or intractable. Particularly for motion planning algorithms, it becomes important to efficiently quantify the uncertainty associated with several admissible trajectories. A sketch of such algorithm is described in Sec.~\ref{KUKAKinEx} (Eq.~(\ref{SigSOptim})). Hence, we propose an alternative probabilistic model where the Jacobian matrix is perturbed by a zero-mean Gaussian noise matrix. The covariance of the noise matrix is obtained through solving an optimization problem to maximize differential entropy of pdf of the noise function. Given an upper bound on the norm of the Jacobian matrix, a constraint is incorporated into the optimization problem that adaptively leverages the uncertainty depending on the dexterity of the manipulator. 

Random Jacobian matrix is written as
\begin{eqnarray}
&& \ \ \ \ \ \ \bm{J}_{k}=\underline{J}_{k}+{\bm{J}_{\nu}}_{k} \label{GausMod}\\
&& \Rightarrow \ \bm{J}_{k}\sim N_{n,n}(\underline{J}_{k},I\otimes\Sigma_{k}) \nonumber
\end{eqnarray}
\noindent
where ${\bm{J}_{\nu}}_{k}\sim N_{n,n}(\bm{0},I\otimes\Sigma_{k})$ and $I\otimes\Sigma_{k}$ implies that rows of the Jacobian matrix are independent random vectors \cite{Gupta}. 
Now, let us define $\bm{Y}={\bm{J}_{\nu}}_{k}^{T}{\bm{J}_{\nu}}_{k}$. Using Theorems 1 and 3, $\bm{Y}\sim W_{n}(n,\Sigma)$. The objective is to find $\Sigma$ such that the entropy of the $p_{\bm{Y}}(Y)$ is maximized. It can be shown \cite{bishop} that maximizing the entropy $S(p_{\bm{Y}}(Y))$ is equivalent to maximizing $f=\ln{|\Sigma|}$. Let us now assume that the upper bound on the Frobenius norm of the $\bm{J}_{k}$ is known and denoted as $u$, {\it i.e.}, $\|\bm{J}\|_{F}\le u$. We showed that \cite{SoviziASME} this constraint along with the model described by Eq.~(\ref{GausMod}) leads to the inequality $\mbox{tr}\big\{n\Sigma\big\}\le u^{2}-\|\underline{J}_{k}\|_{F}^{2}$. Further, to control the level of uncertainty we introduce the parameter $\alpha$ and modify the inequality into $\mbox{tr}\big\{n\Sigma_{k}\big\}\le \alpha^{2}(u^{2}-\|\underline{J}_{k}\|_{F}^{2})$. Finally, the matrix optimization problem is
\begin{empheq}[]{align}
& \mbox{Maximize} \ \ f=\ln{|\Sigma_{k}|}\label{cost}\\
      & \mbox{Subject to}\nonumber\\
			& \mbox{tr}\big\{n\Sigma_{k}\big\}\le \alpha^{2}(u^{2}-\|\underline{J}_{k}\|_{F}^{2}) \label{const1}\\
			& \Sigma_{k}\ge 0 \label{const2}
\end{empheq}

Algorithm \ref{AlgGauss} summarizes the Monte Carlo simulation based on the Gaussian Jacobian matrix model. 
 
\begin{algorithm}
    \caption{Monte Carlo simulation based on Gaussian Jacobian matrix model. \textbf{Inputs}: $q_{k}$, $n$ and $u$; \textbf{Output}: $\Sigma_{k}$, ${J}_{k}$ and ${q}_{k+1}$} \label{AlgGauss}
    \begin{algorithmic}[1]
		  \STATE Substitute $q_{k}$ in the $J(q_{k})$ to obtain $\underline{J}_{k}$
			\STATE Find $\|\underline{J}_{k}\|_{F}$
			\STATE Obtain $\Sigma_{k}$ by solving (\ref{cost})-(\ref{const2})
			\STATE Given $\underline{J}_{k}$ and $\Sigma_{k}$, generate ${J}_{k}$ from matrix-variate normal distribution (MATLAB's command \verb+mvnrnd+ may be useful here)
			\STATE Substitute ${J}_{k}$ in Eq.~(\ref{RMTInvDiff}) to obtain a sample of $\bm{q}_{k+1}$
    \end{algorithmic}
\end{algorithm}

In the experimental study, we show that modeling the inverse of the Jacobian matrix as random matrix (and then following the same procedure described in this subsection) provides smoother uncertainty bounds. Adopting this model in $\dot{\bm{q}}_{k}=\bm{J}_{k}^{-1}\dot{x}_{k}^{d}$ facilitates integrating new uncertainty-quantifying terms (closed-form expressions are tractable due to the Gaussianity) into the motion planning algorithms that are commonly used to generate optimal trajectories for the robotic manipulators. Although motion planning is not the focus of this paper, we briefly describe the procedure and provide a sketch of the optimization setup in Sec.~\ref{PartFiltExp}.      
\subsection{Wrench uncertainty analysis}
\label{GaussProd}
In this section, a systematic treatment to the wrench uncertainty in multi-agent cooperative systems is presented. Uncertainty in the output wrench is mainly induced by configuration/calibration and actuation/sensing uncertainties. Neglecting the effect of these variations in such loosely interconnected systems may result in lack of stability or failure in several applications. However, characterization of the uncertainty introduced by each agent and formulating the resultant effect are practically and theoretically difficult and in some cases intractable. We show that random matrix theories can be used to systematically model the effect of these uncertainties with significantly less characterization efforts. Since limited available information is used in formulating the RMT-based model, it may provide lower fidelity compared to the parametric approaches in which all random entities are carefully modeled (see \cite{SoviziTRO}). However, constructing these parametric models is not feasible in several complex systems and one may seek alternative approaches to appropriately model system variations. We will show that such (RMT-based) approaches can still represent the real system behavior with sufficient accuracy and become particularly more beneficial when the complexity of the system increases ({\it e.g.}, cooperative systems with large number of agents).     

Using the parametric formulation \cite{SoviziTRO}, we estimate the parameters of the RMT-based model when only some bounds on the parameters are given. Through an exhaustive evaluation we show that RMT-based model provide acceptable accuracy as far as system parameters remain in the known bounds. 
\subsubsection{RMT-based formulation}
\label{RMTWrench}
Wrench vector in a system with $m$ agents and $n$ degrees of freedom at the end-effector can be written as
\begin{equation}
\bm{W}=\bm{S}\bm{T}
\label{wrench}
\end{equation} 
\noindent
where $\bm{W}\in \mathbb{R}^{n}$ is random wrench vector, $\bm{S} \in \mathbb{M}_{n\times m}$ is random (static) Jacobian matrix and $\bm{T} \in \mathbb{R}^{m}$ is random force vector. Randomness of the Jacobian matrix and force vector are due to the configuration/calibration and actuation/sensing uncertainties, respectively. 

Let us assume ${\bm{S}}\sim \mathcal{N}_{n,m}(\underline{S},\Sigma_{\bm{S}}\otimes \Psi_{\bm{S}})$ and $\bm{T}\sim \mathcal{N}_{m}(\underline{T},\Sigma_{\bm{T}})$. Further, let us assume $\bm{S}$ and $\bm{T}$ are independent. As Gaussian models can properly represent the behavior of many physical systems, one appropriate assumption is to consider random actuation vector $\bm{T}$ to have a multivariate normal distribution. The Gaussianity assumption on $\bm{S}$ matrix is made by neglecting the deviation of the actual distribution of this matrix from the matrix-variate normal distribution. This facilitates subsequent analytical treatments and the error induced by this approximation can remain bounded an sufficiently low in several cases depending on the system complexity. Moreover, assuming that in a physical system actuators performance is independent of their arrangement, the $\bm{S}$ and $\bm{T}$ independence assumption can be justified. Now, the covariance of the random wrench vector $\bm{W}$ can be derived as follows:
\begin{align}
& \mbox{Cov}(\bm{W})=E\big[\bm{W}\bm{W}^{T}\big]-E\big[\bm{W}\big]E\big[\bm{W}\big]^{T}\nonumber\\
&=E\big[\bm{S}\bm{T}{\bm{T}}^{T}\bm{S}^{T}\big]-E\big[\bm{S}\big]E\big[\bm{T}\big]E\big[\bm{T}\big]^{T}
E\big[\bm{S}\big]^{T}
\label{TorqCov}
\end{align}

The first term on the right hand side of Eq.~(\ref{TorqCov}) can be written as 
\begin{align}
& E\big[\bm{S}\bm{T}{\bm{T}}^{T}\bm{S}^{T}\big]=\int_{\mathcal{D}_{\bm{S}}}
\int_{\mathcal{D}_{\bm{T}}}\bm{S}{\mathbfcal{{T}}}\bm{S}^{T}p_{\bm{S},\bm{T}}(S,T)\mbox{d}\bm{T}\mbox{d}\bm{S}\nonumber \\
&=\int_{\mathcal{D}_{\bm{S}}}
\bm{S}\bigg[\int_{\mathcal{D}_{\bm{T}}}\mathbfcal{{T}}p_{\bm{T}}(T)\mbox{d}\bm{T}\bigg]\bm{S}^{T}p_{\bm{S}}(S)\mbox{d}\bm{S}\nonumber\\
&=E\bigg[\bm{S}E\big[\mathbfcal{{T}}\big]\bm{S}^{T}\bigg]=
\mbox{tr}\big\{E\big[\mathbfcal{{T}}\big]^{T}\Psi_{\bm{S}}\big\}\Sigma_{\bm{S}}
+\underline{S}E\big[\mathbfcal{{T}}\big]\underline{S}^{T}
\label{TorqCovRHS1}
\end{align}
\noindent
where $\mathbfcal{{T}}=\bm{T}{\bm{T}}^{T}$. The last equality in Eq.~(\ref{TorqCovRHS1}) results from Theorem 5. Furthermore, we have 
\begin{equation}
E\big[\mathbfcal{{T}}\big]=\Sigma_{{T}}+\underline{T} \
{\underline{T}}^{T}
\label{MeanOfDynParTrns}
\end{equation}

Substituting Eq.~(\ref{MeanOfDynParTrns}) into (\ref{TorqCovRHS1}), we get
\begin{align}
E\big[\bm{S}\bm{T}{\bm{T}}^{T}\bm{S}^{T}\big]&=\mbox{tr}\bigg\{\bigg[\Sigma_{{T}}+\underline{T} \
{\underline{T}}^{T}\bigg]\Psi_{\bm{S}}\bigg\}\Sigma_{\bm{S}}\nonumber\\
&+\underline{S}\big[\Sigma_{{T}}+\underline{T} \
{\underline{T}}^{T}\big]\underline{S}^{T}
\label{RHS2}
\end{align}

Finally, substituting Eq.~(\ref{RHS2}) into Eq.~(\ref{TorqCov}) gives 
\begin{equation}
\mbox{Cov}(\bm{W})=\mbox{tr}\bigg\{\bigg[\Sigma_{{T}}+\underline{T} \ 
{\underline{T}}^{T}\bigg]\Psi_{\bm{S}}\bigg\}\Sigma_{\bm{S}}
+\underline{S}\Sigma_{{T}}\underline{S}^{T}
\label{CovWNonPar}
\end{equation}    
\subsubsection{Parametric formulation}
More elaborate formulation of the wrench covariance can be developed when the behavior of all agents are statistically identifiable. This implies that, sources of uncertainty are known in some reasonable details and information on the distributions of the random variables are available.      

\begin{figure}[htbp!]
\centering
\includegraphics[trim=100 100 100 30, clip,scale=0.35]{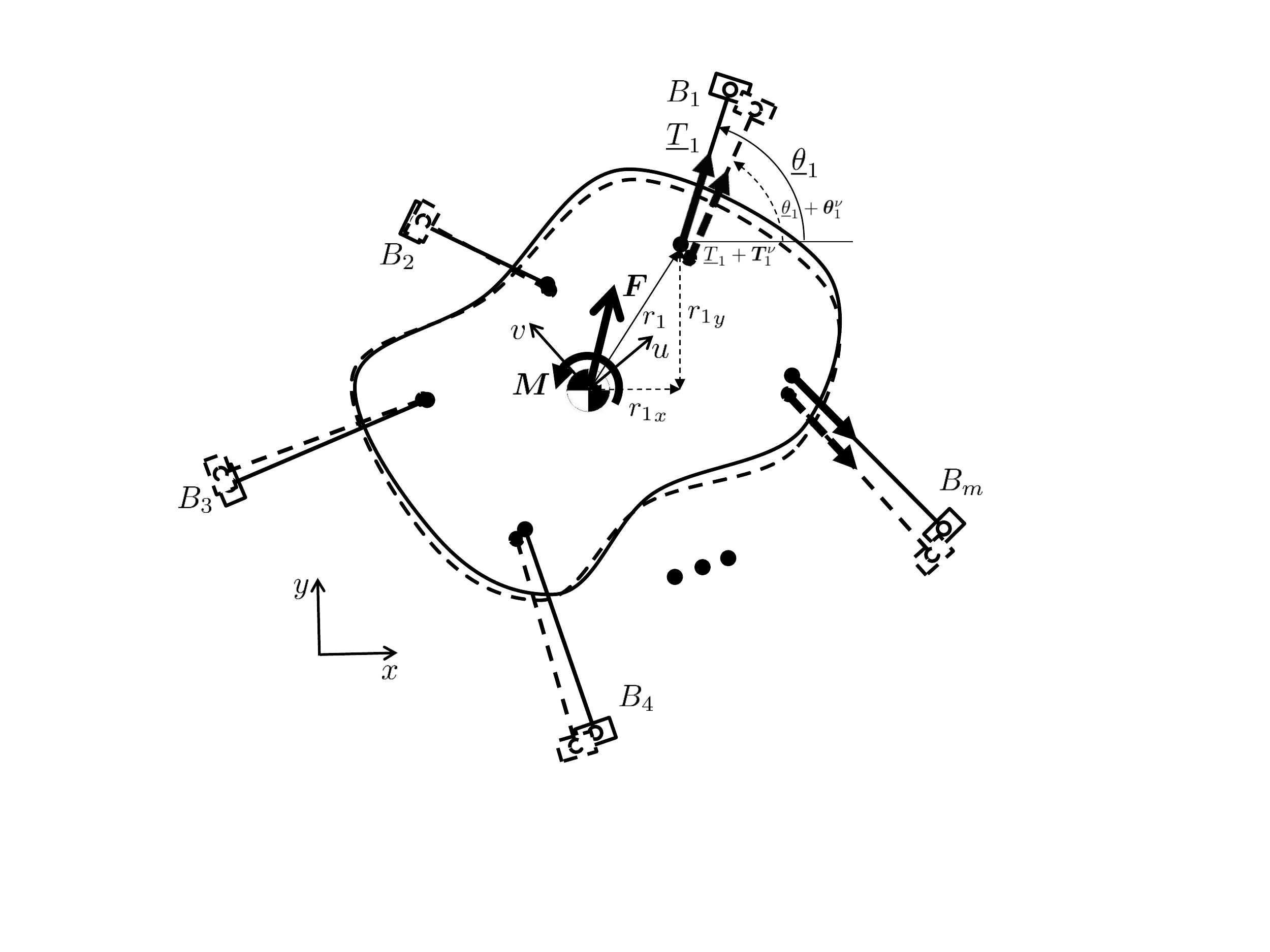}       
\caption{Schematic of a planar cable robot. Solid lines show the nominal (mean) state of the system (tension and orientation of the cables) and dashed lines show the perturbed state.} 
\label{MotivFig}
\end{figure}

Configuration and actuation uncertainties are two major sources of the uncertainty that impact the performance of the multi-agent cooperative systems. Hence, the random total wrench at the platform can be written as the sum of individual wrenches as (a planar system is considered here) 
{\small
\begin{align}
&{\bm{W}}=\begin{bmatrix}
{\bm{F}}_{x}\\
{\bm{F}}_{y}\\
\bm{M}
\end{bmatrix}=\sum_{i=1}^{m}\begin{bmatrix}
{{\bm{f}}_{i}}_{x}\\
{{\bm{f}}_{i}}_{y}\\
\bm{m}_{i}
\end{bmatrix}=\sum_{i=1}^{m}\begin{bmatrix}
\bm{T}_{i}\cos(\bm{\theta}_{i})\\
\bm{T}_{i}\sin(\bm{\theta}_{i})\\
{r_{i}}_{x}{{\bm{f}}_{i}}_{y}-{r_{i}}_{y}{{\bm{f}}_{i}}_{x}
\end{bmatrix}
\label{WrenchStatEq}
\end{align}   
}%
where $\bm{\theta}_{i}=\underline{\theta}_{i}+{\bm{\theta}}_{i}^{\nu}$ and $\bm{T}_{i}=\underline{T}_{i}+{\bm{T}}_{i}^{\nu}$ are the random orientation and tension of the $i${th} agent force vector in which ${\bm{\theta}}_{i}^{\nu}$ and ${\bm{T}}_{i}^{\nu}$ are the perturbation terms.  In Eq.~(\ref{WrenchStatEq}), $r_{i}={r_{i}}_{x}\hat{e}_{1}+{r_{i}}_{y}\hat{e}_{2}$ is the vector in global frame $x-y$ pointing from the origin of the local frame $u-v$ to the attachment point of the $i${th} cable, as shown in Fig.~\ref{MotivFig}. 

We assume that the magnitudes of the force vectors ($\bm{T}_{i}$'s) have Gaussian distributions with mean $\underline{T}_{i}$ and standard deviation $\sigma_{T_{i}}$, {\it i.e.}, $\bm{T}_{i}\sim \mathcal{N}(\underline{T}_{i},\sigma_{T_{i}}^{2})$. Moreover, we use von Mises distribution, that is in fact a counterpart of the Gaussian distribution on the circle, in order to model random orientation of the force vectors. Random variable $\bm{\theta}$ has von Mises distribution with mean $\underline{\theta}$ and dispersion parameter $\sigma_{\theta}$, denoted by $\bm{\theta}\sim \mbox{vMF}(\underline{\theta},\sigma_{\theta})$ if its pdf is given by
\begin{equation}
p_{{\bm{\theta}}}({\theta})=\frac{\displaystyle 1}{\displaystyle 2\pi I_{0}({\sigma}_{\theta})}\mbox{exp}\{{\sigma}_{\theta}\cos({\theta}-\underline{\theta})\}, \ {\bm{\theta}}\in [0,2\pi]
\label{vMis}
\end{equation}

So, we have $\bm{\theta}_{i}\sim \mbox{vMF}(\underline{\theta}_{i},\sigma_{\theta_{i}}) \ \forall i=1,\dots,m$. We also incorporate the following independence assumptions: (i) $\bm{\theta}_{i}$ and $\bm{\theta}_{j}$ are independent $\forall i,j=1,\dots,m$ where $i\neq j$; (ii) $\bm{T}_{i}$ and $\bm{T}_{j}$ are independent $\forall i,j=1,\dots,m$ where $i\neq j$; and (iii) $\bm{T}_{i}$ and $\bm{\theta}_{j}$ are independent $\forall i,j=1,\dots,m$. We showed that the variances of the wrench vector elements can then be calculated as \cite{SoviziTRO, sovizi2016application}
\begin{eqnarray}
&& \mbox{Var}(\bm{F}_{x})=\sum_{i=1}^{m}\bar{a}_{i}\cos^{2}(\underline{\theta}_{i})-\bar{b}_{i}\cos(2\underline{\theta}_{i})\label{kapF11}\\
&& \mbox{Var}(\bm{F}_{y})=\sum_{i=1}^{m}\bar{a}_{i}\sin^{2}(\underline{\theta}_{i})+\bar{b}_{i}\cos(2\underline{\theta}_{i}) \label{kapF22}\\
&&  \bar{a}_{i}=\underline{T}_{i}^{2}+\sigma_{T_{i}}^{2}-\left(\frac{\displaystyle I_{1}(\sigma_{\theta_{i}})}{\displaystyle I_{0}(\sigma_{\theta_{i}})}\underline{T}_{i}\right)^{2}\nonumber\\
&& \bar{b}_{i}=\frac{\displaystyle I_{1}(\sigma_{\theta_{i}})\left(\displaystyle \underline{T}_{i}^{2}+\sigma_{T_{i}}^{2}\right)}{\displaystyle I_{0}(\sigma_{\theta_{i}})\sigma_{\theta_{i}}} \nonumber
\end{eqnarray}
\noindent
where $I_{n}(z)=\frac{\displaystyle 1}{\displaystyle \pi}\displaystyle \int_{0}^{\pi}\mbox{exp}\{z\cos(x)\}\cos(nx)\mbox{d}x$ is the $n$th order (when $n$ is integer) modified Bessel function of the first kind. We use the closed-form expressions of the force variance given by Eqs.~(\ref{kapF11}) and (\ref{kapF22}) to estimate the parameters of the model given by Eq.~\ref{CovWNonPar}, when only some bounds on the system parameters are known. The parameter estimation scheme is described in Sec.~\ref{NumStudWr} in the context of a numerical example.
\subsubsection{Numerical study}
\label{NumStudWr}
We are aiming to optimally find the parameters of the RMT-based model described in Eq.~(\ref{CovWNonPar}) such that the model captures the behavior of the systems whose parameters fall into the specific known bounds. Following information are known in our numerical simulations: (i) upper and lower bounds on $\sigma_{\theta_{i}}$'s {\it i.e.}, $\sigma_{\theta}^{l}\le \mbox{min}\{\sigma_{\theta_{1}},\dots,\sigma_{\theta_{m}}\}$ and $\sigma_{\theta}^{u}\ge \mbox{max}\{\sigma_{\theta_{1}},\dots,\sigma_{\theta_{m}}\}$; (ii) $\sigma_{T_{i}}$'s, the lower bound $\sigma_{T}^{l}\le \mbox{min}\{\sigma_{T_{1}},\dots,\sigma_{T_{m}}\}$ and the upper bound $\sigma_{T}^{u}\ge \mbox{max}\{\sigma_{T_{1}},\dots,\sigma_{T_{m}}\}$; and (iii) the domain on which the system mean state is defined, {\it i.e.}, $\underline{T}_{i}^{l}\le \underline{T}_{i} \le \underline{T}_{i}^{u}$ and $\underline{\theta}_{i}^{l}\le \underline{\theta}_{i} \le \underline{\theta}_{i}^{u}$ ($\forall i=1,\dots,m$).  

Now, let us consider the output force in a planar cable-based system in which all cables are attached to one point on the platform (zero moment). The static Jacobian matrix ${S}$ is a $2\times m$ matrix whose first and second rows are the cosine's and sine's of the force vector orientations, {\it i.e.},
\begin{equation}
\bm{S}=\begin{bmatrix}
\cos(\bm{\theta}_{1}), \dots, \cos(\bm{\theta}_{m})\\
\sin(\bm{\theta}_{1}), \dots, \sin(\bm{\theta}_{m})\\
\end{bmatrix}
\end{equation}

Relying on the independence of $\bm{\theta}_{i}$'s, it is a valid assumption to consider the columns of $\bm{S}$ as independent random vectors, hence, in Eq.~(\ref{CovWNonPar}) $\Psi_{\bm{S}}=I_{m}$ \cite{Gupta}. So, $\Sigma_{S}$ is the only parameter that needs to be optimally chosen given the lower and upper bounds on the system state and parameters.  

In 1000 Monte Carlo simulations, a system sample is first obtained by drawing a sample of $\sigma_{\theta_{i}}$, $\sigma_{T_{i}}$, $\underline{T}_{i}$ and $\underline{\theta}_{i}$ from uniform distributions with corresponding (given) bounds. In our numerical simulation we set $\sigma_{\theta}^{l}=100$, $\sigma_{\theta}^{u}=800$, $\sigma_{T}^{l}=0.5 \ \mbox{N}$, $\sigma_{T}^{u}=1 \ \mbox{N}$ and the bounds on the mean tension and orientation of the cables are $\underline{T}_{i}^{l}=3 \ \mbox{N}$, $\underline{T}_{i}^{u}=5 \ \mbox{N}$, $\underline{\theta}_{i}^{l}=\pi/4 \ \mbox{rad}$ and $\underline{\theta}_{i}^{u}=\pi \ \mbox{rad}$ ($\forall i=1,\dots,m$). Then, the following optimization problem is solved to obtain the optimal covariance matrix $\Sigma_{\bm{S}}$:
{\small
\begin{align}
\label{SigSOptim}
\underset{\Sigma_{\bm{S}}\ge 0}{\mbox{Minimize}} \ \ F&=\|diag(\mbox{Cov}(\bm{W}))-[\mbox{Var}(\bm{F}_{x}), \ \mbox{Var}(\bm{F}_{y})]^{T}\| 
\end{align}}%
\noindent
in which $\mbox{Cov}(\bm{W})$ is given by Eq.~(\ref{CovWNonPar}) and $\mbox{Var}(\bm{F}_{x})$ and $\mbox{Var}(\bm{F}_{y})$ are given by Eq.~(\ref{kapF11}) and Eq.~(\ref{kapF22}), respectively. Finally, we set the optimal value of $\Sigma_{\bm{S}}$ to be the average of 1000 $\Sigma_{\bm{S}}$'s obtained in Monte Carlo runs. Note that the optimization problem described in (\ref{SigSOptim}) targets only the diagonal terms of $\Sigma_{\bm{S}}$. This is because only the distance between variances of the force vector are enforced to be minimized and the covariance ($\mbox{Cov}(\bm{F}_{X},\bm{F}_{Y})$) terms are not included in the objective function. The optimal values of the off-diagonal terms can be found by slight modification of (\ref{SigSOptim}) where the objective function is augmented with the distance between the covariance terms corresponding to the parametric and RMT-based models. This can further improve the accuracy of the RMT-based model results.
\begin{figure}[b!]
	\centering
	\includegraphics[trim=0 0 0 10, clip, scale=0.85]{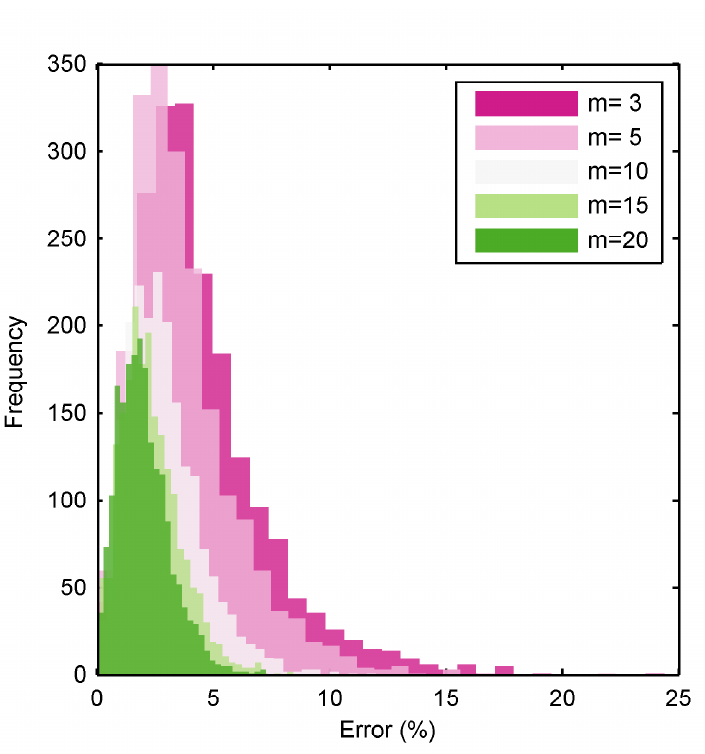}
	\caption{Error histograms. For 2000 samples of system state and parameters, the Euclidean distances (error) between the output wrench variances obtained by RMT-based model and those given by parametric model are calculated. 5 histograms with different colors correspond to 5 systems with different number of cables.}
	\label{ErrWrnRMT}
\end{figure}

Figure~\ref{ErrWrnRMT} shows the error histograms when the RMT-based model is tested for 2000 random system realizations. Note that the system samples are different from those (1000 samples) used for parameter estimation. As shown in Fig.~\ref{ErrWrnRMT}, for a simple system with only $m=3$ agents and in 2000 trials, the errors remain less than 25 \% and the average error is 4 \% which is significantly lower than the maximum error. The notable fact is that by increasing the number of agents, error histograms are continuously shifting toward lower values. For example, maximum error for a system with $m=20$ agents is 7 \% and the error average is 2 \%. This indicates a significant improvement of the RMT-based model performance when compared to the results corresponding to a system with only 3 agents. These results encourage replacing the detailed parametric models by RMT-based models especially when the number of uncertain agents grows in complex multi-agent systems. However, a more general conclusion can be drawn after careful theoretical analysis that can be the subject of the future work. In Sec.~\ref{WrenchExper}, we examine the performance of the RMT-based model developed in this section, in capturing the output wrench of a real experimental setup consisting of three mobile robots (iRobot).  

\section{Experimental studies}
\label{ExpRes}
Here, we experimentally validate RMT-based approaches proposed in the preceding sections. For motion uncertainty analysis a KUKA youBot arm is utilized. Moreover, for validation of the RMT-based model developed for wrench uncertainty quantification, we use multiple iRobots that are providing a static wrench at the platform.           
\subsection{Motion uncertainty analysis: KUKA youBot}
\label{KUKAKinEx}
Random Jacobian matrix approach, developed in Sec.~\ref{KinAnaly}, is investigated here using a KUKA youBot arm shown in Fig.~\ref{KUKAyouBot}. The goal of our experiment on youBot is twofold: (i) to investigate the performance of the RMT-based models and if they provide higher fidelities and outperform the conventional approaches; and (ii) to provide more accurate bounds on the joint states estimation error when RMT-based kinematic motion model is used coupled with the low-cost Kinect sensor measurements. 
\subsubsection{Experimental setup} 
\label{KukExpSetup}
\begin{figure}[htb!]
\includegraphics[trim=175 100 150 70, clip, scale=0.6]{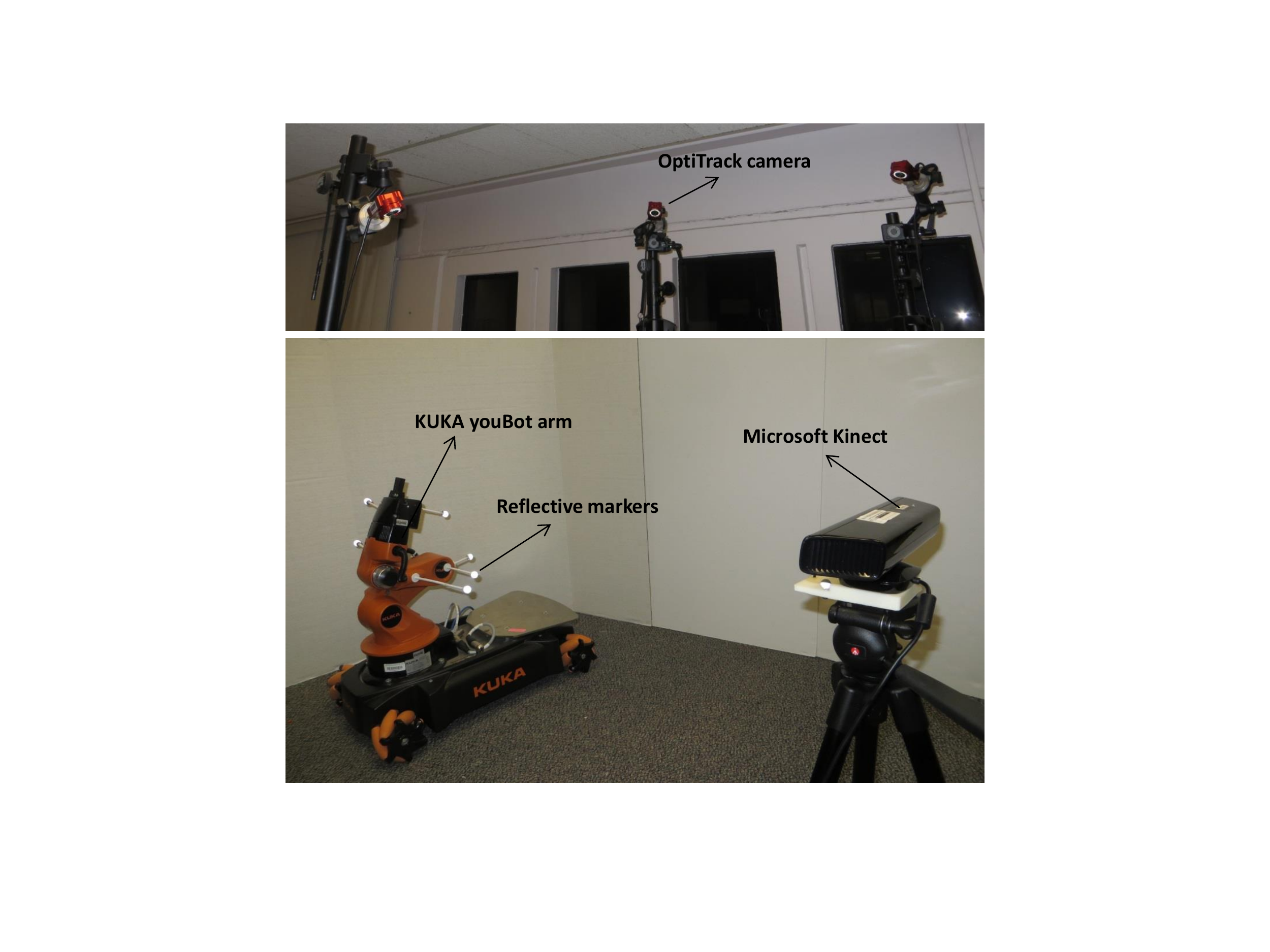}
\caption{Experimental setup for motion uncertainty analysis: KUKA youBot with reflective markers on its three links; Kinect sensor to observe robot configuration; and OptiTrack cameras for motion capturing (ground truth data).}
\label{KUKAyouBot}
\end{figure}

We use marker-based motion capturing system in order to provide (approximately) ground truth values of the joint angles at each time instant. Three reflective markers are attached to each link of the youBot arm and OptiTrack cameras are used to capture the trajectory of the markers. Note that youBot arm has 5 joints (excluding the end-effector gripper) from which two of them are set to be fixed during the experiment: first joint that is located at the base and rotates the entire arm; and  the joint located at the wrist. This makes the system a three-links serial chain. We use low-cost Kinect sensor as our sensing modality. The experimental setup is shown in Fig.~\ref{KUKAyouBot}.     
\subsubsection{Motion uncertainty characterization}
\begin{figure}[htb!]
\includegraphics[trim=0 0 0 0, clip, scale=1]{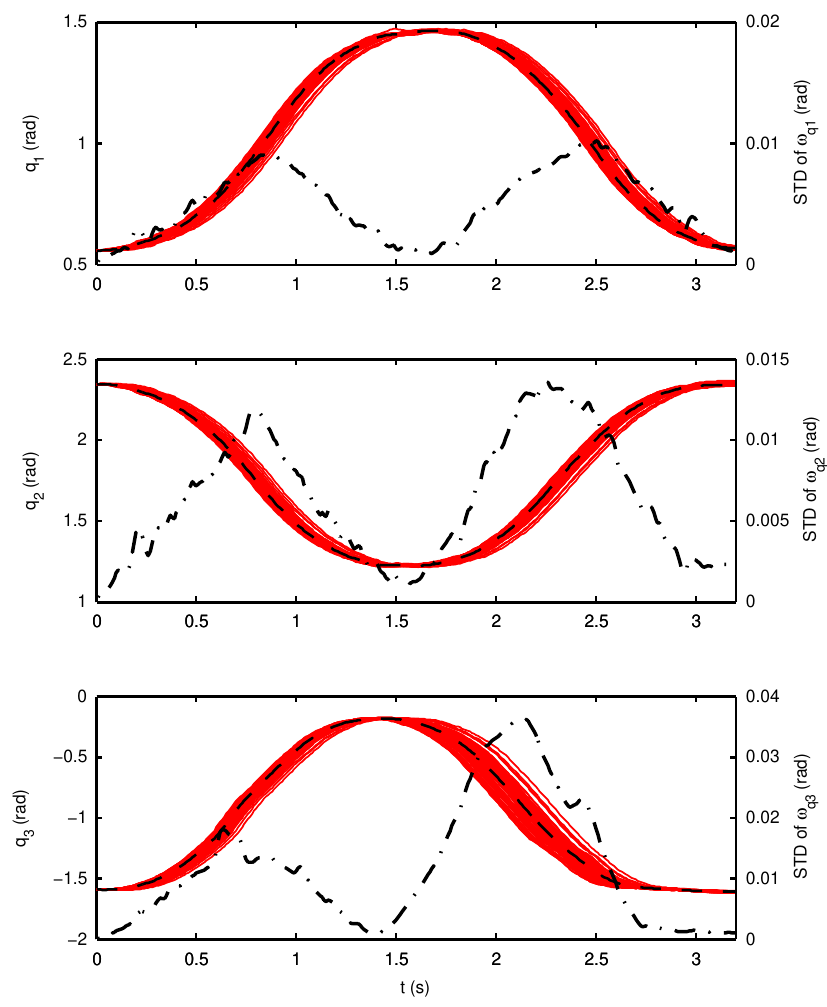}
\caption{Left ordinate: realizations (solid red lines) of the KUKA youBot joint angles in a trajectory tracking task performed in $T=3.2$ seconds (for each realization). Dashed black lines are the mean of each ensemble overlaid on top of the realizations. Right ordinate: standard deviation of the process noise (shown by dashdot lines) obtained from experimental data and using Eq.~(\ref{OmgReal}).}
\label{JointRealzFig}
\end{figure}

A desired end-effector trajectory is first defined. Then, we repeat the experiment (trajectory tracking) for 100 times while the desired trajectory remains unchanged. This provides us with 100 realizations of the joint angles of the robot arm captured through relatively accurate motion capturing system. The data set is used as ground truth for our further statistical analysis. Realizations of the joint angles along with their corresponding mean trajectories are shown in Fig.~\ref{JointRealzFig}.     

Right ordinate in Fig.~\ref{JointRealzFig} shows the sample estimate of the standard deviation of the process noise $\bm{\omega}_{k+1}$ in Eq.~(\ref{InvDiffRndm}), calculated from experimental data as follows. Let us denote the true (measured) state of the system at $k$th time instant in $i$th realization by $\tilde{q}_{k}^{i}$. Now, the $i$th realization of the process noise can be obtained as 
\begin{equation}
\omega_{k+1}^{i}=q_{k+1}^{i}-\tilde{q}_{k+1}^{i}=\tilde{q}_{k}^{i}+J(\tilde{q}_{k}^{i})^{-1}\dot{x}_{k}^{d}\Delta t-\tilde{q}_{k+1}^{i}
\label{OmgReal}
\end{equation} 

A visual inspection of Fig.~\ref{JointRealzFig} shows the dynamic nature of the uncertainty level in the process noise $\bm{\omega}_{k+1}$ in Eq.~(\ref{InvDiffRndm}). Now, we examine three stochastic models in capturing the uncertainty in the system state depicted in Fig.~\ref{JointRealzFig}: (i) an inverse differential kinematics model perturbed by an additive Gaussian noise; (ii) an RMT-based model with Wishart perturbation matrix (using Algorithm \ref{WishMCAlg}); and (iii) an RMT-based model with Gaussian noise matrix (using Algorithm \ref{AlgGauss}). \cite{karydis2015} developed a general methodology to evaluate the fidelity of stochastic models corresponding to different robotic platforms. Adopting their terminology and notation, a model $\mathcal{M}$ to describe the stochastic behavior of a system can be parameterized by $p$-dimensional vector $\xi\in \mathbb{R}^{p}$. So, once the structure of the model is proposed, the problem turns into finding the best $\xi$ such that $\mathcal{M}(\xi)$ captures the system response most accurately. Here, for additive Gaussian noise model, the parameter vector is $\xi=[\sigma_{1,1}^{\bm{\omega}},\sigma_{1,2}^{\bm{\omega}}, \dots, \sigma_{3,3}^{\bm{\omega}}]$ that is in fact the elements of the noise covariance matrix. For RMT-based model with Wishart perturbation matrix, we have $\xi=\sigma_{\bm{B}}$ that is the scalar dispersion parameter and with Gaussian noise matrix the model parameter vector is $\xi=[\alpha,u]$. 
 
Once a set of experimental measurements are available, different strategies can be used to estimate the parameter vector $\xi$. For example in \cite{karydis2015}, a lease-squares method is proposed to minimize the distance between model output and average of the experimentally-obtained samples at each time instant. Here, we first calculate the samples of process noise at each time point using Eq.~(\ref{OmgReal}). Then, for additive Gaussian noise model, the covariance matrix of the noise is considered to be time-invariant and equal to the mean of the sample estimates of covariance matrices at each time instant, {\it i.e.}, $\hat{\Sigma}_{\bm{\omega}}=(1/M)\sum_{k=1}^{M}\hat{\Sigma}_{k}$. Capturing for 3.2 seconds with 100 frames per second (fps) sampling frequency results in $M=320$. For RMT-based model with Wishart perturbation matrix, we calculate $\sigma_{\bm{B}}$ as follows. First, from the model proposed in Eq.~(\ref{RMTInvDiff}), the process noise added to the state at $t_{k+1}$ is
\begin{equation}
\bm{\omega}_{k+1}=(\bm{J}^{-1}_{k}-\underline{J}^{-1}_{k})\dot{x}^{d}_{k}\Delta t
\label{Omegk}
\end{equation}

Hence, given $\tilde{q}_{k}^{i}$ and $\omega_{k+1}^{i}$ from Eq.~(\ref{OmgReal}) and substituting $\bm{J}_{k}$ from Eq.~(\ref{LUDecom}), we have
\begin{equation}
(({\underline{J}_{1}^{i}}_{k}{B}_{k+1}^{i}{\underline{J}_{2}^{i}}_{k})^{-1}-{\underline{J}_{k}^{i}}^{-1})\dot{x}^{d}_{k}\Delta t={\omega}_{k+1}^{i}
\label{Omegki}
\end{equation}         
\noindent
where $\underline{J}_{k}^{i}=J(\tilde{q}_{k}^{i})$. Then, $i$th sample of the perturbation matrix $\bm{B}$ at time $t_{k+1}$ ({\it i.e.}, $B_{k+1}^{i}$) can be obtained by solving the following optimization problem. 
\begin{align}
\underset{{B_{k+1}^{i}}^{-1}}{\mbox{Minimize}} \ \ F&=\alpha_{1}\|{\omega}_{k+1}^{i}-({\underline{J}_{2}^{i}}_{k}^{-1}{{B}_{k+1}^{i}}^{-1}{\underline{J}_{1}^{i}}_{k}^{-1}\nonumber\\
&-{\underline{J}_{k}^{i}}^{-1})\dot{x}^{d}_{k}\Delta t\|+\alpha_{2}\|{B_{k+1}^{i}}^{-1}-I\|\label{BsampOpt}\\
& \mbox{s.t.}\nonumber\\
& {B_{k+1}^{i}}^{-1}> 0\label{const10}
\end{align}%
\noindent
where $\alpha_{1}+\alpha_{2}=1$. Assuming the inverse of $B_{k+1}^{i}$ rather than $B_{k+1}^{i}$ as optimization variable facilitates using the convex optimization packages, and here, we use \verb+CVX+ \cite{CVX2,CVX1} for solving the problem described by (\ref{BsampOpt})-(\ref{const10}). Last term in the objective function enforces $B_{k+1}^{i}$ to be close to its mean, {\it i.e.}, $I_{n}$. Using the samples of $\bm{B}_{k}$ (at $t_{k}$), $\sigma_{\bm{B}_{k}}$ can be calculated using Eq.~(\ref{NormStd}). Finally, we calculate the dispersion parameter as $\hat{\sigma}_{\bm{B}}=(1/M)\sum_{k=1}^{M}\hat{\sigma}_{\bm{B}_{k}}$. In RMT-based model with Gaussian noise matrix, we set
\begin{equation}
\hat{u}=\underset{k=1,\dots,M}{\mbox{max}}\{\|\underline{J}_{k}^{-1}\|\}+\tilde{u}
\end{equation}
\noindent
where $\tilde{u}$ is a positive constant that extends the deterministic upper bound on Jacobian norm due to the uncertainty. Note that for this model, we formulate the inverse of the Jacobian matrix as a random matrix because it provides smoother bounds on the uncertainty level. Here, we choose $\tilde{u}$ and $\alpha$ such that (in 100 realizations) the mean and variance of the model outputs take the minimum distance from those of actual experiments, {\it i.e.}, 
\begin{align}
\underset{\tilde{u},\alpha}{\mbox{Minimize}} \ \ F&=\sum_{k=1}^{M}\beta_{1}\|\underline{q}_{k}-\underline{\tilde{q}}_{k}\|\nonumber\\
&+\beta_{2}\mbox{tr}\{\mbox{Cov}({q}_{k})-\mbox{Cov}({\tilde{q}}_{k})\}
\end{align}
\noindent
where $\beta_{1}+\beta_{2}=1$. The results of our parameter estimations for three different models are
{\small
\begin{align*}
& \mbox{Add Gauss:} \ \ \ \ \hat{\Sigma}_{\bm{\omega}}=\begin{bmatrix}
    0.0330 &  -0.0406  &  0.0592\\
   -0.0406 &   0.0547  &  -0.0839\\
    0.0592 &  -0.0839  &  0.2457\\
\end{bmatrix}\times 10^{-3}\\
& \mbox{RMT- Wish:} \ \ \ \hat{\sigma}_{\bm{B}}=0.25\\
& \mbox{RMT- Gauss:}\ \ \hat{u}=18, \  \hat{\alpha}=0.1
\end{align*}}%

\begin{figure}[htb!]
\centering
\includegraphics[trim=0 0 0 0, clip, scale=0.85]{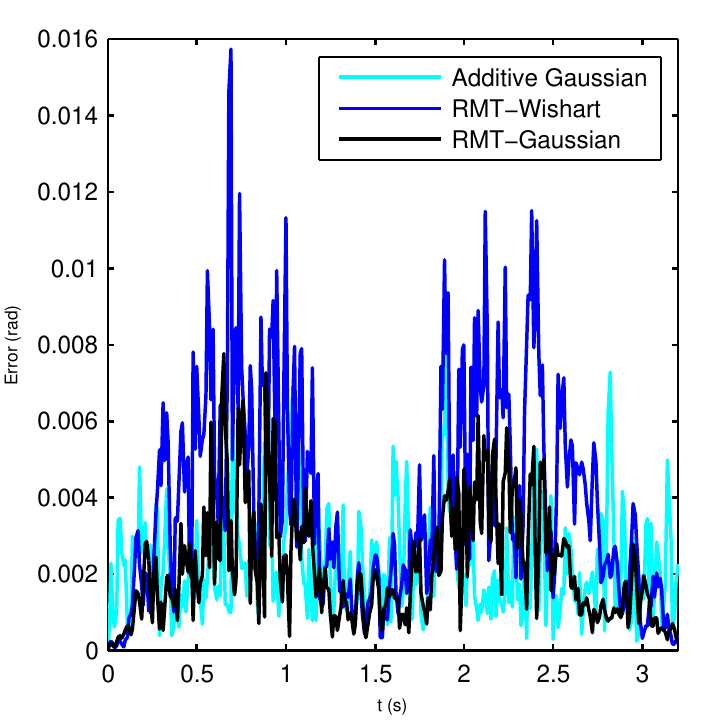}
\caption{Joint states error when the mean of model-generated realizations is compared with the mean of real experimental observations. All three models can appropriately capture the mean in this example where the maximum error is less than 0.02 rad.}
\label{MeanError}
\end{figure}

Given the models parameters, we generated 100 realizations of the joint states for each stochastic model. Joint states at each time instant are perturbed with a sample of Gaussian noise (in MATLAB using \verb+mvrnd+ function) when additive Gaussian noise model is adopted. When using RMT-based model with Wishart perturbation matrix, a sample of Jacobian matrix at each time instant is used to propagate the state that can be drawn using Algorithm~\ref{WishMCAlg}. Finally, when RMT-based model with Gaussian noise matrix is used, at each time step we sample the Jacobian matrix using the procedure described in Algorithm~\ref{AlgGauss}.            
Figure~\ref{MeanError} shows the error corresponding to each model. Error is computed as the Euclidean distance between the mean of joint states generated through model simulations and that of real experimental data at each time instant. 

\begin{figure}[htb!]
\centering
\includegraphics[trim=0 0 0 0, clip, scale=1]{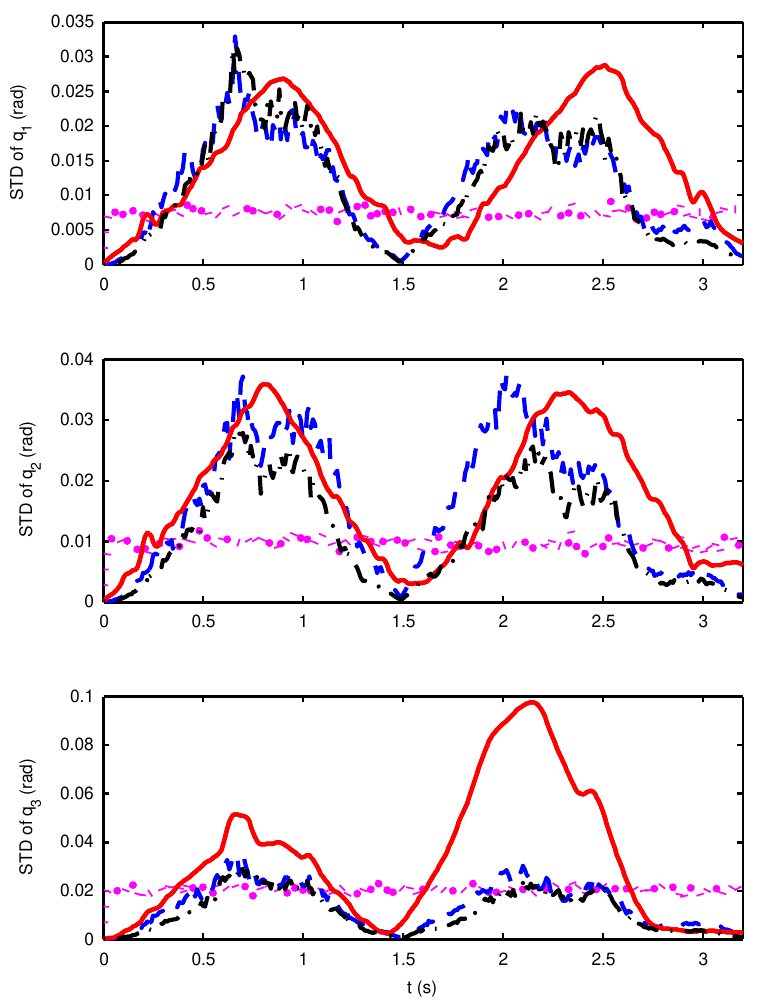}
\caption{Standard deviation of the joint angles obtained from experimental (accurate) observations, and three stochastic models simulations. Red solid line shows the experimental results. The results obtained from additive Gaussian noise model is shown by cyan dotted lines. Blue dashed lines and black dot-dash lines correspond to RMT-based models with Wishart perturbation and Gaussian noise matrix, respectively.}
\label{STDq} \vspace{-0.2in}
\end{figure}

The error corresponding to Wishart-based model shows higher variations than the other two models in an average sense. However, the maximum error for all models is less than $0.02 \ \mbox{rad}$. This implies the fact that all approaches are capable of capturing the system response up to the first statistical moment with an acceptable error. However, the focus is more on the uncertainty level that is characterized with higher level moments. Standard deviations of the joint states corresponding to the experimental data and three stochastic models are shown in Figure~\ref{STDq}.                

Red solid lines in Fig.~\ref{STDq} are standard deviations corresponding to actual experiment, that are in fact identical to those plotted in Fig.~\ref{JointRealzFig} by black dot-dash lines. Standard deviations corresponding to the generated realizations of the joint states using additive Gaussian noise model, Wishart-based model and the model based on Gaussian noise matrix, are shown with cyan dotted lines, blue dashed lines and black dot-dash lines, respectively. From Fig.~\ref{STDq}, it is clear that the additive Gaussian noise model is unable to properly capture the dynamics of the uncertainty level that is inherent to the actual platform. This is in fact the main drawback of such models that motivates RMT-based approaches. While the results from additive Gaussian noise model remain almost time-invariant, it can be seen that the variation in uncertainty level is properly captured by both RMT-based models. There exists an excessive uncertainty in the third joint angle ($\bm{q}_{3}$) that can not be captured as well as other two joints through RMT-based models. We recognized an unusual clearance in the third joint of our KUKA youBot arm and the excessive raise of the uncertainty is believed to be due to this abnormality.               

The fact that RMT-based models provide higher fidelities compared to other conventional approaches (as depicted in Fig.~\ref{STDq}) motivates application of these models to practical estimation and control problems when sequential Monte Carlo filtering approaches are used. More accurate bounds on the state estimation errors are expected when the conventional models are replaced by RMT-based models. We now examine this through an experiment where RMT-based models are used coupled with Kinect measurements to estimate the state and the bounds on the estimation errors.      

\subsubsection{RMT-based particle filtering}
\label{PartFiltExp}
Sequential Monte Carlo sampling techniques rely on the samples of random variables rather their joint pdf in order to approximate the moments of the posterior. This enables these approaches to be used for state estimation in nonlinear systems (such as the case of our study here) where conventional Kalman filters can not be employed. Here, our purpose is to integrate the RMT-based sampling approaches with a sequential importance sampling filter that estimates the joint angles of the KUKA youBot arm when low-cost Kinect measurements are available (see Fig.~\ref{KUKAyouBot}). For this purpose, we run the experiment described in Sec.~\ref{KukExpSetup} for 50 more trials and simultaneously collect the joint angles data using OptiTrack motion capturing (as our ground truth) and Kinect sensor (as our sensing modality). For our filtering purpose, we use the same model parameters obtained from our earlier experiment with 100 realizations. In some sense, the set of data obtained from the first experiment with 100 realizations is our training data set and 50 new realizations construct our testing data set.

Kinect sensor can stream depth data with a  sampling rate of $30 \ \mbox{fps}$. In each frame, a 2.5-dimensional data structure is provided that contains the information on the depth of each pixel in the Kinect field of view. This information can be used to track moving objects in the scene. In case of an articulated system such as KUKA youBot arm, one sensor can be used to simultaneously track all the links. This feature along with their low cost makes such sensors desirable for many applications. However, on the other hand, the low quality measurement limits their applications when high-accuracy perceptions of the system state are required. There are commercial software tools (such as nuiCapture) that, in addition to Kinect data acquisition, provide skeleton (articulation) detection and tracking. However, these tracking platforms are mainly built for human motions. For a custom articulated system, {\it i.e.}, KUKA youBot arm, a customized algorithm is required to detect and track the articulated arm from the point cloud. In computer vision literature, there exist several such algorithms that can solve similar problems. However, here we provide a simple algorithm based on the least-squares method that proves to be adequate for our specific system and application. In order to make the paper as self-contained as possible, we briefly discuss this procedure here.

\begin{figure}[b!]
	\centering
	\begin{minipage}[t]{0.22\textwidth}
		\includegraphics[trim=0 0 0 0, clip, scale=1]{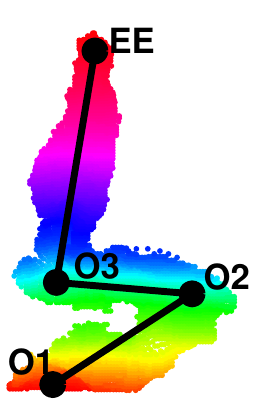}
		
		\hspace{0.35in}(a)
	\end{minipage}%
	\hspace{-0.4in}
	\begin{minipage}[t]{0.22\textwidth}
		\includegraphics[trim=0 0 0 0, clip, scale=1]{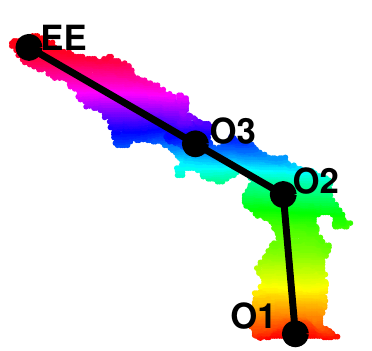} 
		
		\hspace{0.6in}(b)
	\end{minipage}%
	
	\caption{Estimation of the joint angles of the KUKA youBot using Kinect measurement. (a) is the initial and final configuration and (b) shows the most extended configuration while tracking the desired trajectory. In order to improve the estimation of the joint angles, we allowed the length of the segments to change in a certain range.}
	\label{KinOptFig}
\end{figure}

\begin{figure}
	\centering
	\begin{minipage}[t]{0.25\textwidth}
		\includegraphics[trim=0 0 0 0, clip, scale=1]{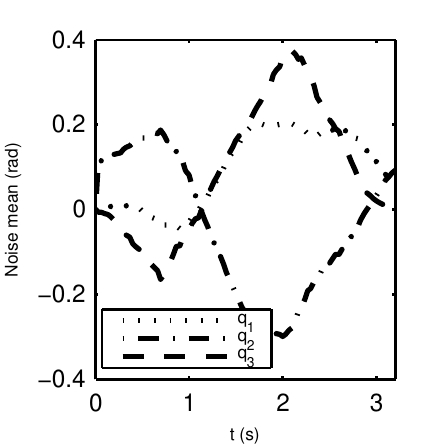}
		
		\hspace{0.9in}(a)
	\end{minipage}%
	\begin{minipage}[t]{0.25\textwidth}
		\includegraphics[trim=0 0 0 0, clip, scale=1]{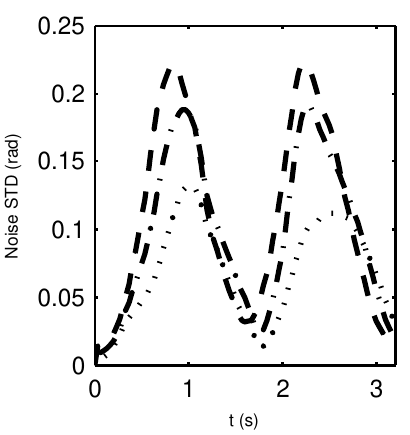} 
		
		\hspace{0.9in}(b)
	\end{minipage}%
	\caption{Noise characteristics of the Kinect sensor.
		(a) shows the mean and (b) shows the standard deviation. It can be seen that Kinect noise is biased and its characteristics are time-varying.} \vspace{-0.2in}
	\label{MeanStdKinNoise}
\end{figure}

We use our prior knowledge of the articulation in KUKA youBot arm in order to detect and track the robot links. In each frame, finding the points corresponding to joint 1 (shown by O1 in Fig.~\ref{KinOptFig}) and the end-effector (shown by EE) is straightforward. This is because O1 position does not change during the robot motion and relative position of the point EE respect to other points is known (from prior information about the end-effector trajectory). Now, in order to fit three segments (corresponding to three links) to the point cloud, we need to find the optimal position of joint 2 and 3, {\it i.e.}, points O2 and O3 in Fig.~\ref{KinOptFig}, respectively. Hence, assuming the point cloud representing the youBot arm is described by $\mathcal{P}=\{p_{1},\dots,p_{I}\}$, we solve the following optimization problem to complete the articulation detection.
\begin{align}
\underset{x_{1},y_{1},x_{2},y_{2}}{\mbox{Minimize}} \ \ F&=\gamma_{1}\bigg[(1/I)\sum_{i=1}^{I} \mbox{min}\{d_{p_{i}}^{s_{1}},d_{p_{i}}^{s_{2}},d_{p_{i}}^{s_{3}}\}\bigg]\nonumber\\
&+\gamma_{2}\bigg[\sum_{j=1}^{3}|d_{a}^{s_{j}}-d_{u}^{s_{j}}|\bigg]
\label{KinObj}
\end{align}%
\noindent
where $(x_{1},y_{1})$ and $(x_{2},y_{2})$ are the coordinates of points O2 and O3, respectively. Note that motion of the robot links is in a plane when the base joint is fixed. In Eq.~(ref{KinObj}), $d_{p_{i}}^{s_{j}}$ is the shortest distance between the $i$th point in the point cloud to the $j$th segment. Moreover, $d_{a}^{s_{j}}$ and $d_{u}^{s_{j}}$ are the maximum distance of the points above and under the $j$th segment. 
First term in (\ref{KinObj}) fits the segments by first finding the point that belong to that segment (link) and minimizing the distance from the points. However, the second term is also necessary to keep the segment passing through the middle of the point cloud corresponding to each link. This is because the density of the points captured by Kinect is different at different locations on the robot link and solely minimizing the distance (first term in the objective function) shifts the segment toward the high density areas, and hence, less symmetry and lower accuracy. The mean and standard deviation of the Kinect noise over the experiment time are shown in Fig.~\ref{MeanStdKinNoise}.      

Now with the Kinect based observations and by using the OptiTrack measurements as the ground truth data, one can examine the performance of three stochastic motion models in joint states estimation. Let us consider the joint states sequence as a Markov process denoted by $\{\bm{q}_{1},\dots,\bm{q}_{M}\}$ ($M=320$) and $\bm{q}_{k}\in \mathcal{D}_{\bm{q}}$ where $\mathcal{D}_{\bm{q}}$ is the domain defined by the robot joint limits. Also, let us denote the Kinect observation at time $t_{k}$ by $\bm{y}_{k}$ and $\bm{y}_{k}$'s are conditionally independent given the state at $t_{k}$. At each time instant $t_{k}$, we seek to estimate the expectations of the posterior defined by
\begin{equation}
E(f_{k}(\bm{q}_{k}))=\int_{\bm{q}_{k}\in \mathcal{D}_{\bm{q}}}f_{k}(\bm{q}_{k})p(\bm{q}_{k}|\bm{y}_{k})\mbox{d}\bm{q}_{k}
\end{equation}

\begin{figure*}[htb!]
	\centering
	{\small \begin{minipage}[t]{0.35\textwidth}
			\center{\qquad $\bm{q}_{1}$}
		\end{minipage}\hfill
		\begin{minipage}[t]{0.325\textwidth}
			\center{\ $\bm{q}_{2}$}
		\end{minipage}\hfill
		\begin{minipage}[t]{0.325\textwidth}
			\center{\ $\bm{q}_{3}$}
	\end{minipage}}\hfill\\
	\begin{minipage}[t]{0.35\textwidth}
		\includegraphics[trim=0 0 0 0, clip, scale=1]{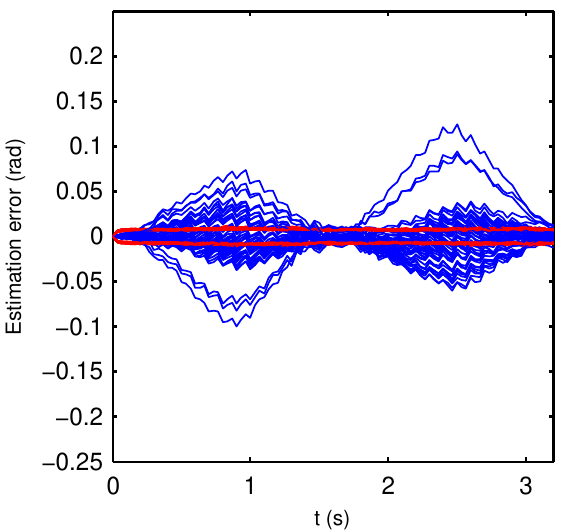}
	\end{minipage}\hfill
	\begin{minipage}[t]{0.325\textwidth}
		\includegraphics[trim=0 0 0 0, clip, scale=1]{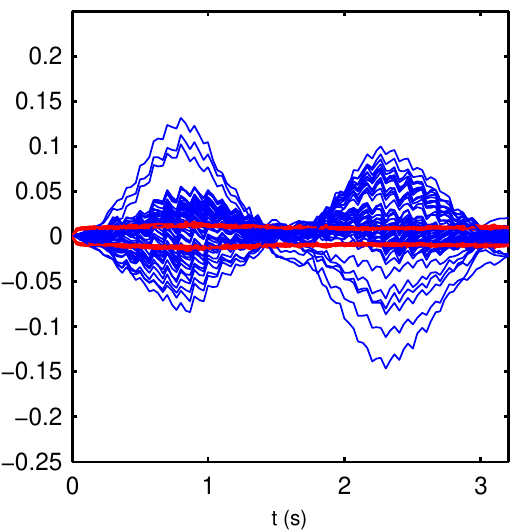} 
	\end{minipage}\hfill
	\begin{minipage}[t]{0.325\textwidth}
		\includegraphics[trim=0 0 0 0, clip, scale=1]{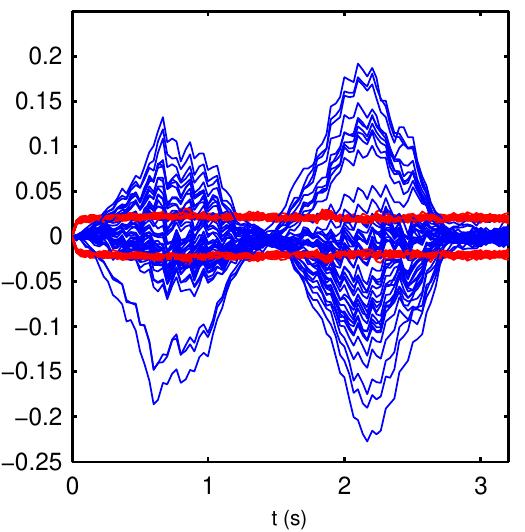} 
	\end{minipage}\hfill\\
	\begin{minipage}[t]{0.35\textwidth}
		\includegraphics[trim=0 0 0 0, clip, scale=1]{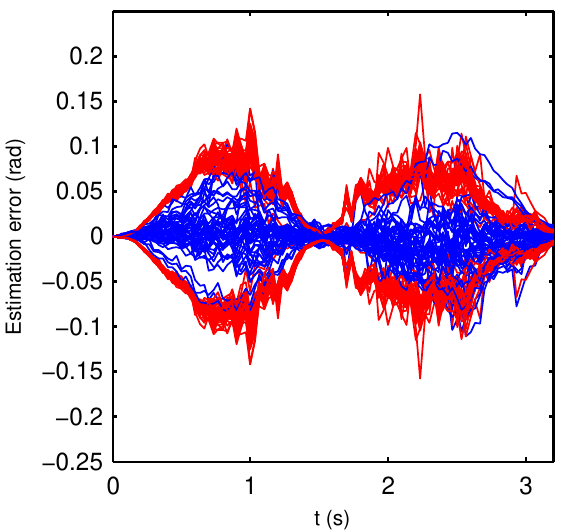}
	\end{minipage}\hfill
	\begin{minipage}[t]{0.325\textwidth}
		\includegraphics[trim=0 0 0 0, clip, scale=1]{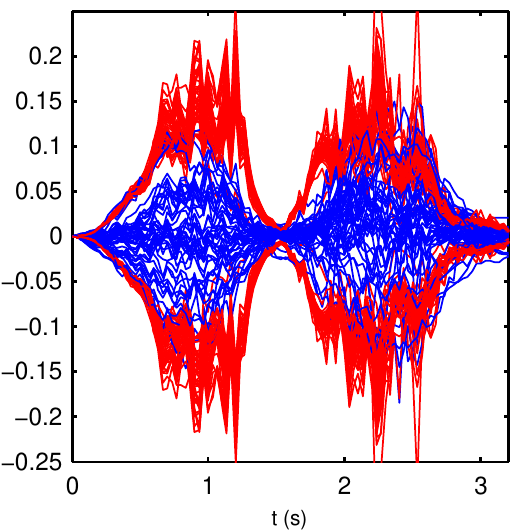} 
	\end{minipage}\hfill
	\begin{minipage}[t]{0.325\textwidth}
		\includegraphics[trim=0 0 0 0, clip, scale=1]{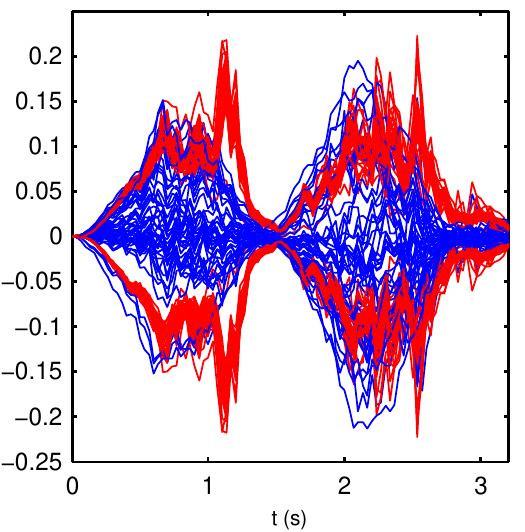} 
	\end{minipage}\hfill\\
	\begin{minipage}[t]{0.35\textwidth}
		\includegraphics[trim=0 0 0 0, clip, scale=1]{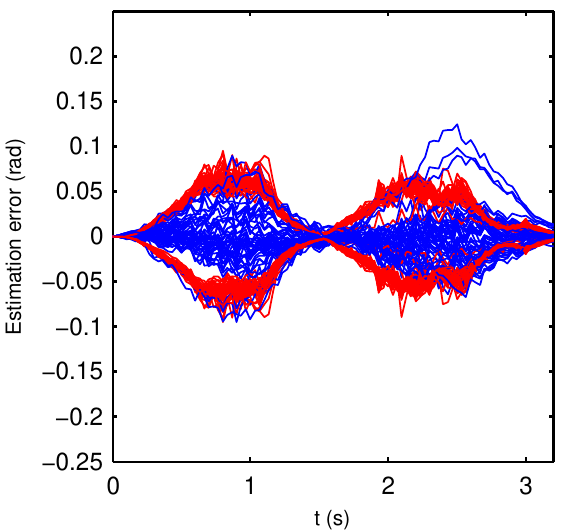}
	\end{minipage}\hfill
	\begin{minipage}[t]{0.325\textwidth}
		\includegraphics[trim=0 0 0 0, clip, scale=1]{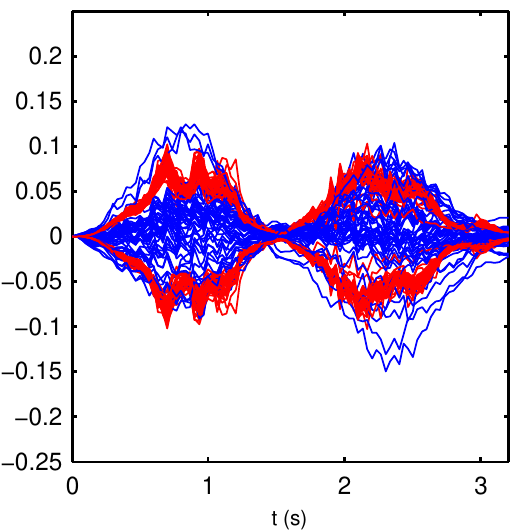} 
	\end{minipage}\hfill
	\begin{minipage}[t]{0.325\textwidth}
		\includegraphics[trim=0 0 0 0, clip, scale=1]{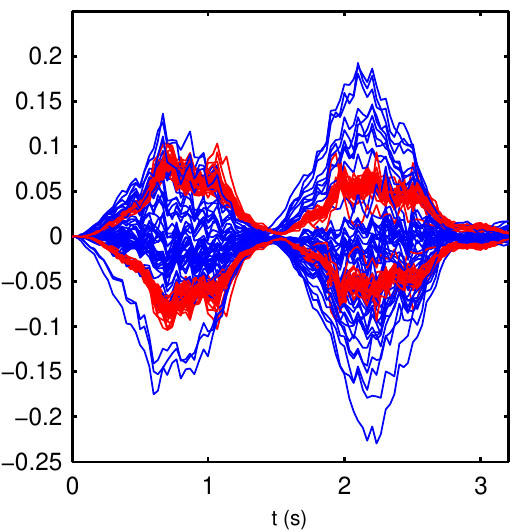} 
	\end{minipage}\hfill\\
	
	\caption{State estimation errors in particle filtering. First row shows the results corresponding to the additive Gaussian noise model. Second row shows the results corresponding to the RMT-based model with Wishart perturbation matrix. Third row corresponds to the RMT-based stochastic model based on Gaussian noise matrix. Red lines are the $1-\sigma$ error bounds and blue lines show the actual error.}
	\label{PartFiltRes} \vspace{-0.2in}
\end{figure*}

We used sequential importance sampling \cite{partfiltDouc} to estimate the mean ($f_{k}(\bm{q}_{k})=\bm{q}_{k}$) and covariance ($f_{k}(\bm{q}_{k})=[\bm{q}_{k}-\underline{q}_{k}][\bm{q}_{k}-\underline{q}_{k}]^{T}$) of the random state $\bm{q}_{k}$ for 50 test trials. The results corresponding to three different models are shown in Fig.~\ref{PartFiltRes}. 
\begin{figure*}
	\centering
	{\small \begin{minipage}[t]{0.27\textwidth}
			\center{Robots configuration\qquad }
		\end{minipage}
		\begin{minipage}[t]{0.22\textwidth}
			\center{Experimental observations}
		\end{minipage}
		\begin{minipage}[t]{0.22\textwidth}
			\center{RMT-based model output  }
	\end{minipage}}\\ \vspace{0.1in}
	\begin{minipage}[t]{0.27\textwidth}
		\includegraphics[trim=0 0 0 0, clip, scale=0.35]{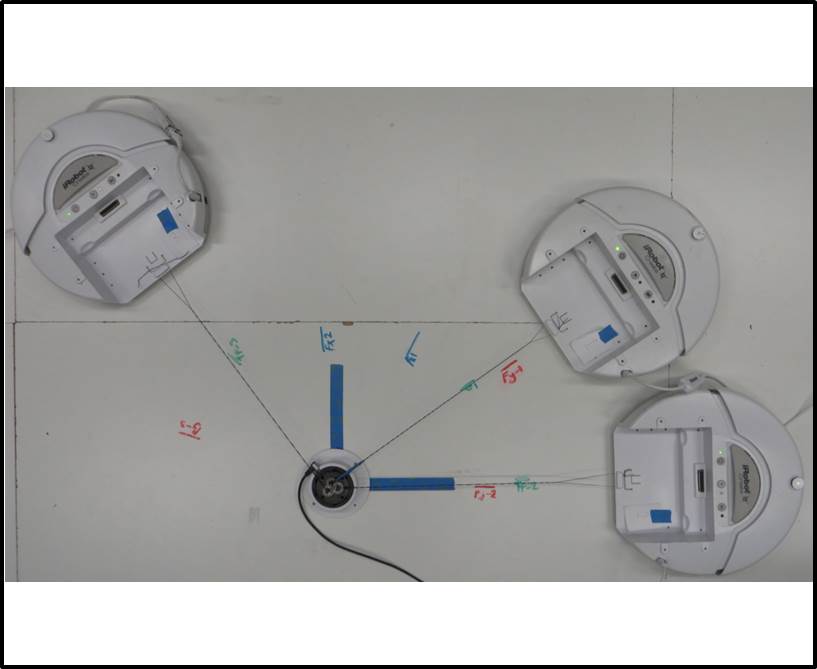}
	\end{minipage}
	\begin{minipage}[t]{0.22\textwidth}
		\includegraphics[trim=80 10 20 0, clip, scale=0.09]{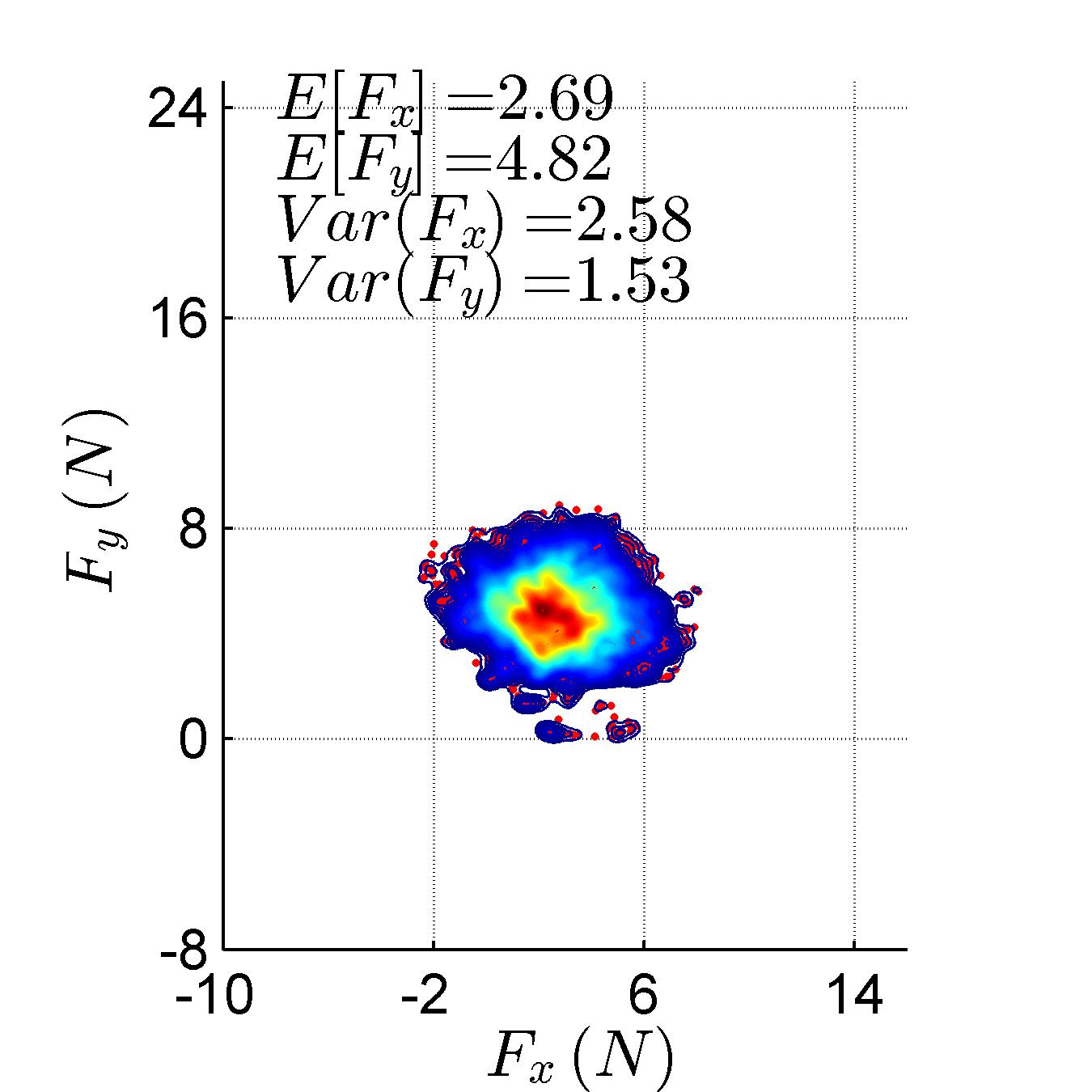} 
	\end{minipage}
	\begin{minipage}[t]{0.22\textwidth}
		\includegraphics[trim=45 10 20 0, clip, scale=0.09]{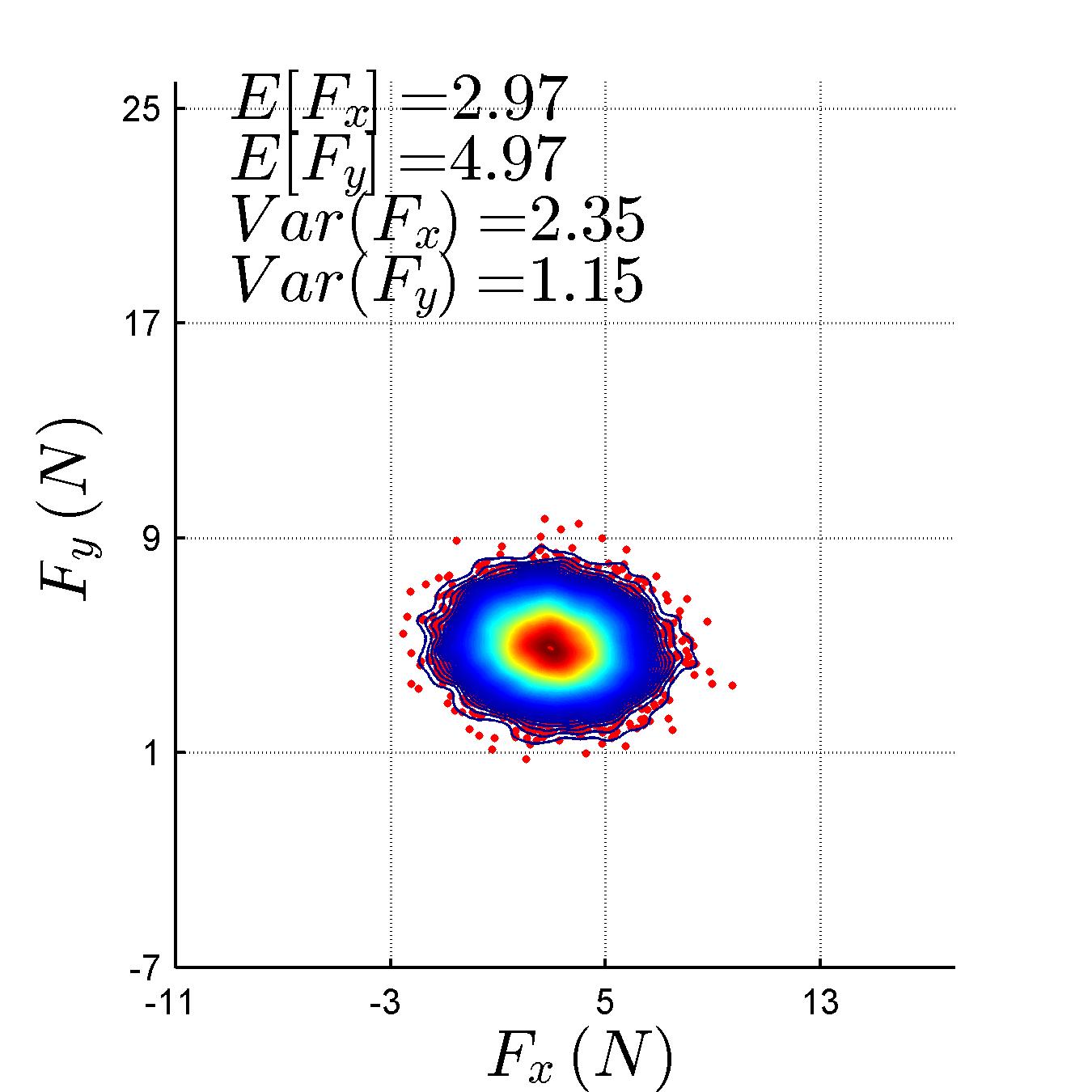} 
	\end{minipage}\\
	\begin{minipage}[t]{0.27\textwidth}
		\includegraphics[trim=0 0 0 0, clip, scale=0.35]{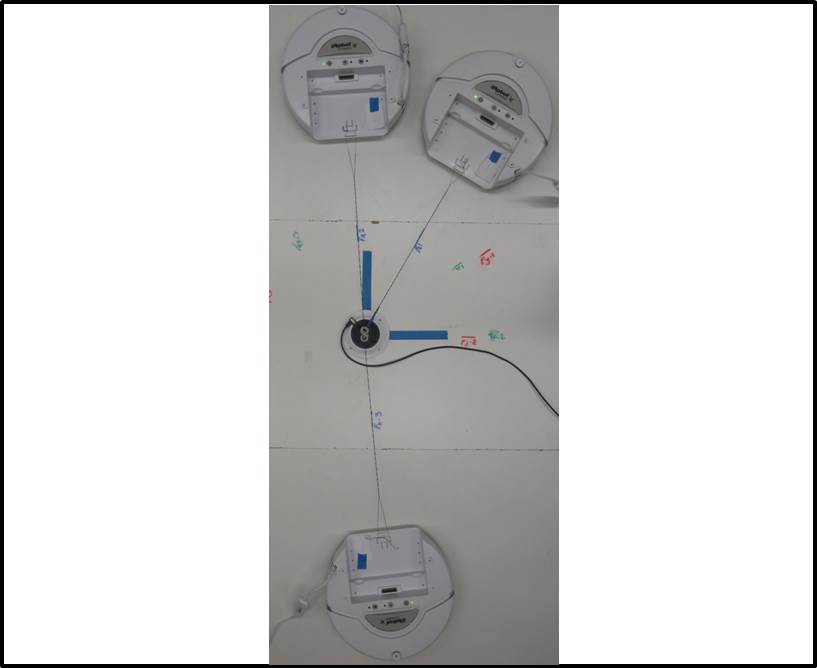}
	\end{minipage}
	\begin{minipage}[t]{0.22\textwidth}
		\includegraphics[trim=155 10 0 0, clip, scale=0.09]{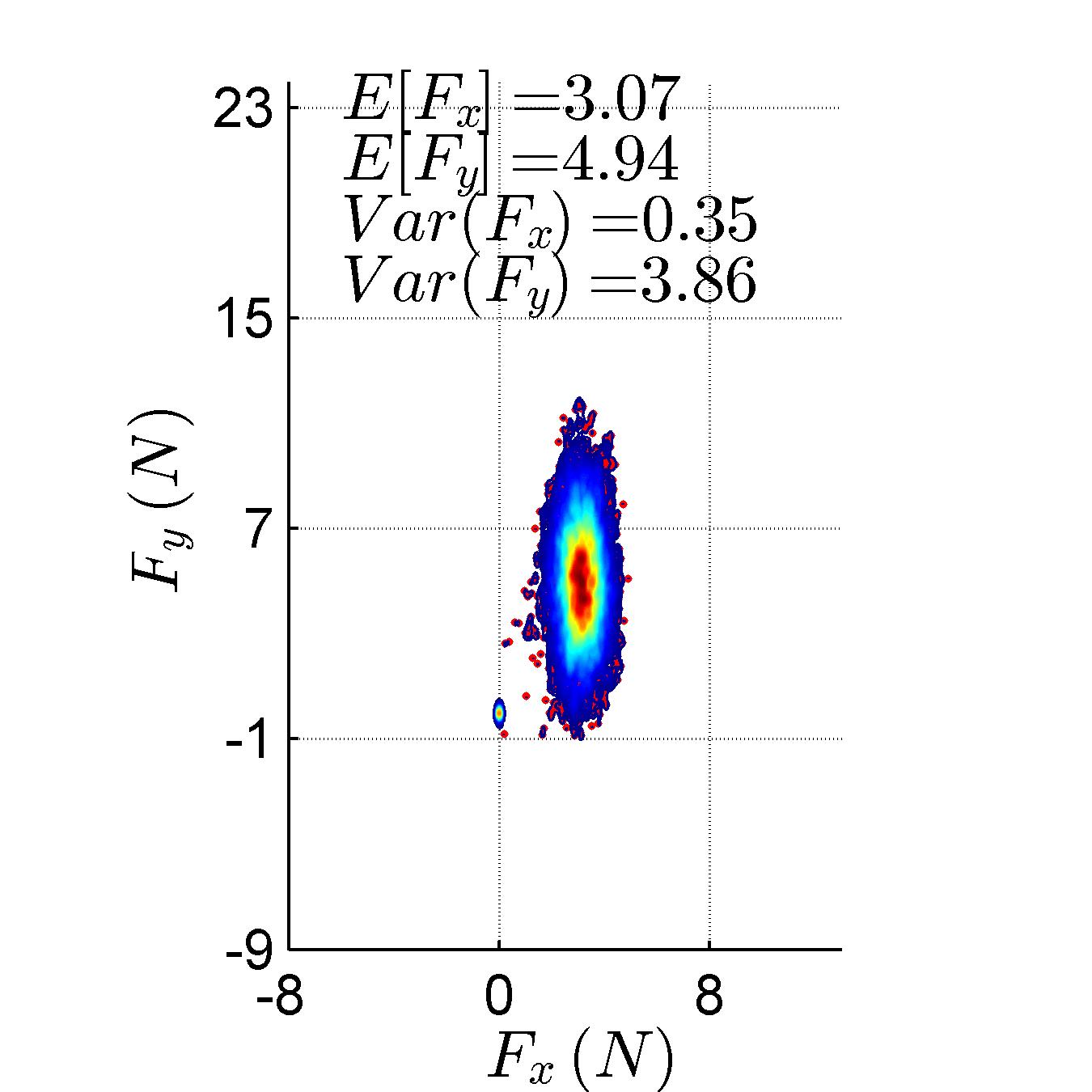} 
	\end{minipage}
	\begin{minipage}[t]{0.22\textwidth}
		\includegraphics[trim=180 10 0 0, clip, scale=0.09]{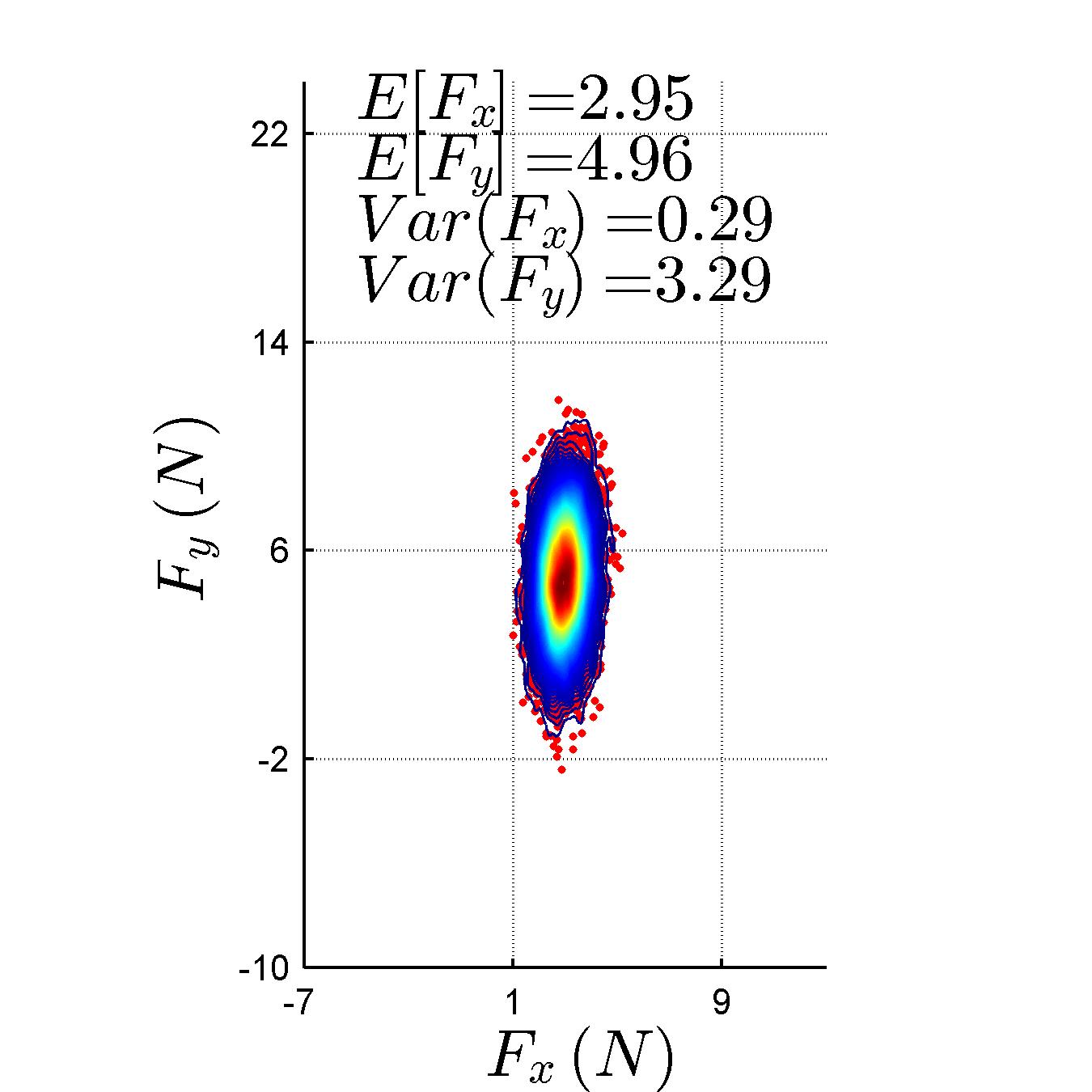} 
	\end{minipage}\\
	\begin{minipage}[t]{0.27\textwidth}
		\includegraphics[trim=0 0 0 0, clip, scale=0.35]{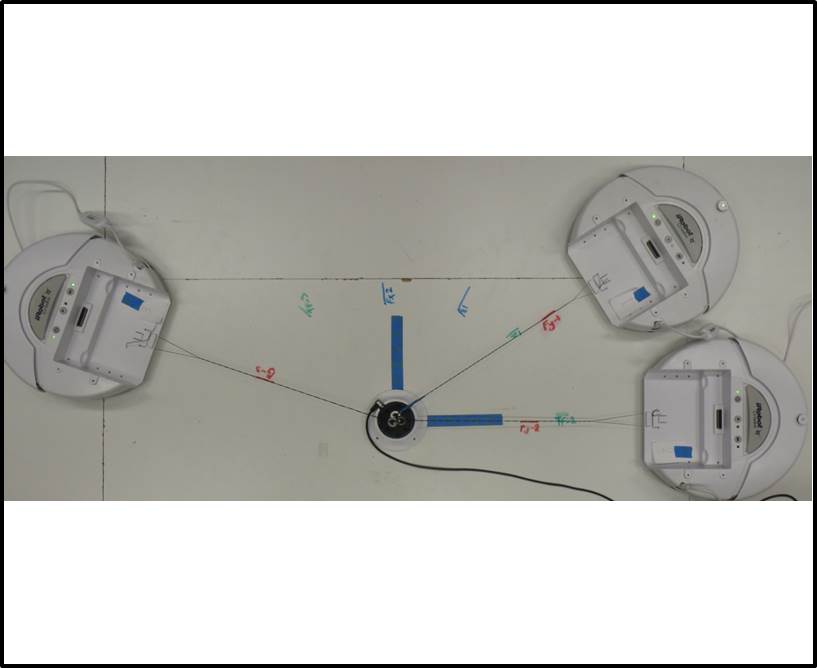}
	\end{minipage}
	\begin{minipage}[t]{0.22\textwidth}
		\includegraphics[trim=0 50 0 0, clip, scale=0.09]{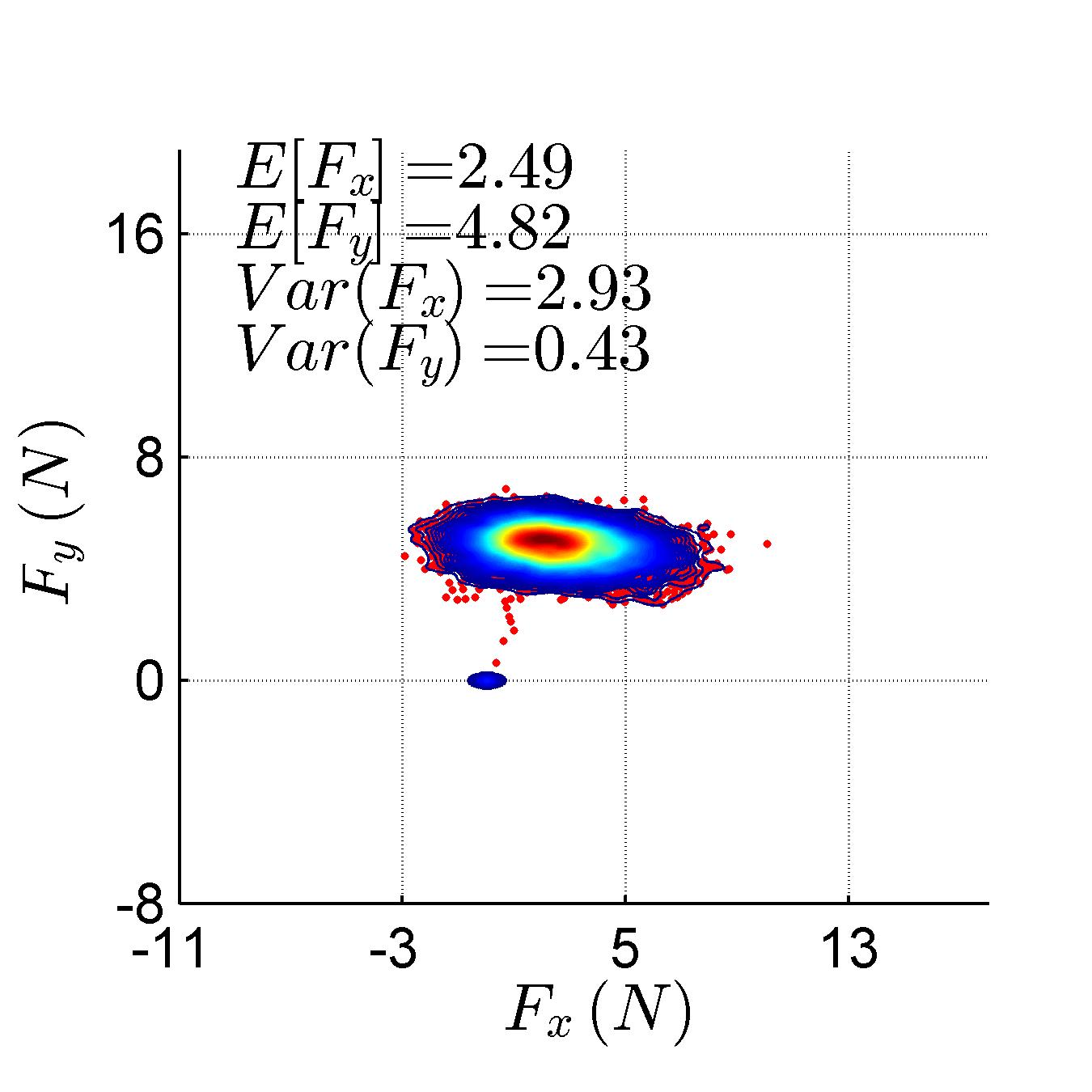} 
	\end{minipage}
	\begin{minipage}[t]{0.22\textwidth}
		\includegraphics[trim=0 50 0 0, clip, scale=0.09]{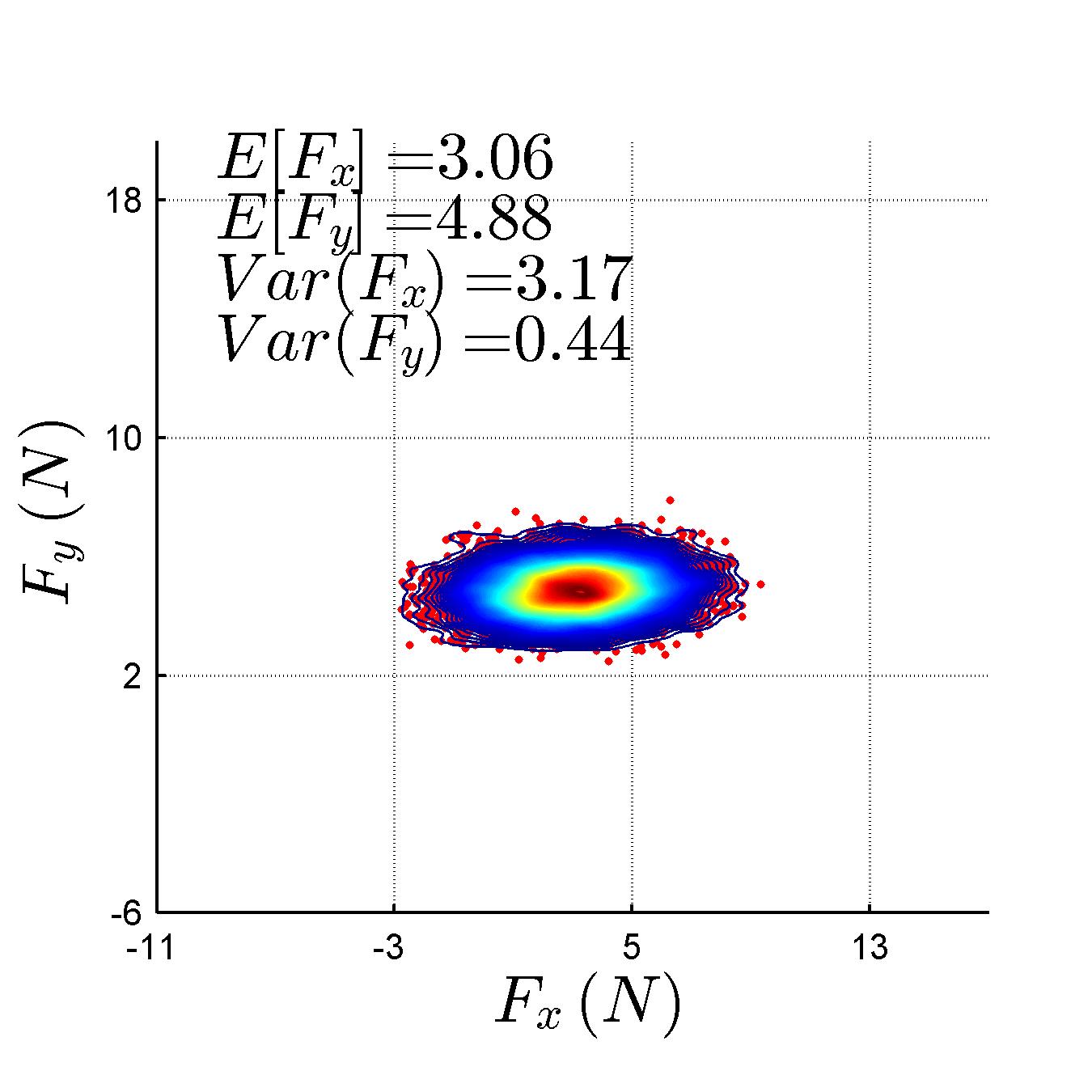} 
	\end{minipage}
	
	\caption{Experimental evaluation of RMT-based model in capturing the wrench (only force vector here) uncertainty. Three different configurations are shown in three rows. First column shows the configuration of the iRobots. The KS density of the (accurately) measured force is shown in second column. The KS density of the data points obtained from RMT-based model given by Eq.~(\ref{CovWNonPar}) is shown in the third column.}
	\label{WrenchExprmnt}\vspace{-0.2in}
\end{figure*}

Blue lines in Fig.~\ref{PartFiltRes} show the actual estimation error that is the distance between the estimated mean and the ground truth value of the state (captured by the OptiTrack cameras). Red lines show the $\pm 1\mbox{-}\sigma$ error bounds on the state estimates. First row of Fig.~\ref{PartFiltRes} corresponds to the additive Gaussian noise model, second row corresponds to the RMT-based model with Wishart perturbation and third row shows the results corresponding to the RMT-based model with Gaussian noise matrix. Comparing the actual error (shown by blue lines) for different models, the results are supporting those that are shown in Fig.~\ref{MeanError} and imply that all three models are capable of capturing the mean state with approximately the same level of accuracy. 

However, comparing the estimates of the error bounds (shown by red lines) shows a significant difference and proves the superiority of the RMT-based models. When additive Gaussian noise model is used (first row of Fig.~\ref{PartFiltRes}), error bounds can not be reliably estimated in the commonly used particle filtering technique. The estimates on the error bounds are approximately constant over the time while the actual error (shown by the blue lines) is highly varying. The estimates of the error bounds are significantly improved and properly capture the variation of the actual error when RMT-based models are used (second and third rows). This shows that both RMT-based stochastic models outperform the commonly used additive Gaussian noise model in estimating the uncertainty level. The bounds corresponding to Wishart-based model are more conservative compared to those from Gaussian noise matrix model. Modeling the inverse of Jacobian matrix in Gaussian RMT-based model results in less conservative results. Despite the small difference, the performance of both the models are relatively close and one may use them interchangeably depending on the particular problem and available information. 

Additionally, the Gaussian RMT-based model can be adopted to augment the motion uncertainty in the motion planning algorithms. Motion planning is not addressed in this paper, however, we briefly discuss the procedure an provide an sketch of the optimization problem as follows. Assume $\bm{q}$ as discrete stochastic process denoted by $\{\bm{q}_{0},\bm{q}_{1},\dots,\bm{q}_{M}\}$ and $\bm{q}_{k}\in \mathcal{D}_{\bm{q}} \ (\forall k=0,\dots,M)$ where $\mathcal{D}_{\bm{q}}$ is the domain on which the joint angles are defined (feasible range of motion of the joints). In order to choose the least uncertain joint trajectory for performing the desired task described by $\{x_{0}^{d};\dot{x}_{0}^{d},\dots,\dot{x}_{M-1}^{d}\}$ one may solve the following optimization problem:%
\vspace{-0.1in}
{\small \begin{align}
\underset{\underline{q}_{0},\underline{q}_{1},\dots,\underline{q}_{M}\in \mathcal{D}_{\bm{q}}}{\mbox{Minimize}} \ \ F&=\sum_{k=1}^{M}\mbox{Cov}(\dot{\bm{q}}_{k})\\
& \mbox{s.t.} \nonumber\\
& \underline{q}_{k+1}-\underline{q}_{k}=J_{k}^{-1}\dot{x}_{k}^{d}  \qquad (\forall k=0,\dots,M-1)\nonumber
\label{SigSOptim}
\end{align}}%
\vspace{-0.1in}

If the Gaussian RMT-based model is adopted, one can consider $\dot{\bm{q}}_{k}=\bm{J}_{k}^{-1}\dot{x}_{k}^{d}$, and hence, the covariance in (\ref{SigSOptim}) has a closed-form expression given in Sec.~\ref{RMTWrench}. Note that this is a rough sketch of the proposed approach where the uncertainty associated with the candidate trajectory is taken into account.  However, practical motion planning algorithms include many factors into the formulation such as collision avoidance, minimization of time and energy/fuel, etc., that are not the focus of this work.
\vspace{-0.3in}
      
\subsection{Wrench uncertainty analysis: cooperative iRobots}
\label{WrenchExper}\vspace{-0.1in}

A system consisting of three iRobots is used to evaluate the RMT-based wrench uncertainty model (Eqs.~(\ref{wrench}) and (\ref{CovWNonPar})). iRobots are tied to an ATI force transducer using cables, as shown in Fig.~\ref{WrenchExprmnt}. A desired output wrench $W^{d}=[3 \ \mbox{N}, \ 5\ \mbox{N}, \ 0 \ \mbox{N.m}]^{T}$ is aimed to be generated at the platform (that is the force transducer here) as a result of the individual tractions provided by each iRobot. Three different configurations of the iRobots, shown in the first column of Fig.~\ref{WrenchExprmnt}, are considered to provide this wrench. Let us refer to the configuration in the first, second and third rows as C1, C2 and C3, respectively. While all configurations provide (approximately) the same mean wrench at the platform, the uncertainty profile is significantly different. See \cite{SoviziTRO} for details on the uncertainty-based optimization of the system configuration. The idea is to investigate the effectiveness of the Gaussian product model in capturing the real system output in a cooperative system. After identifying the characteristics of each iRobot, the bounds on the system parameters are set to be $\sigma_{\theta}^{l}=169.35$, $\sigma_{\theta}^{u}=2.416\times 10^{3}$, $\sigma_{T}^{l}=0.48$, and $\sigma_{T}^{u}=2.27$. The mean state bounds for three different configurations are ($\forall i=1,\dots,3$)  
\begin{align*}
& \mbox{C1:} \qquad \underline{T}_{i}^{l}=1.82 \ \mbox{N}, \ \underline{T}_{i}^{u}=13.89 \ \mbox{N} \nonumber\\
& \quad \ \ \qquad \underline{\theta}_{i}^{l}=0 \ \mbox{rad}, \ \underline{\theta}_{i}^{u}=3.8 \ \mbox{rad} \nonumber\\
& \mbox{C2:} \qquad \underline{T}_{i}^{l}=1.82 \ \mbox{N}, \ \underline{T}_{i}^{u}=13.89 \ \mbox{N} \nonumber\\
& \quad \ \ \qquad \underline{\theta}_{i}^{l}=\pi/4 \ \mbox{rad}, \ \underline{\theta}_{i}^{u}=2\pi \ \mbox{rad} \nonumber\\
& \mbox{C3:} \qquad \underline{T}_{i}^{l}=1.82 \ \mbox{N}, \ \underline{T}_{i}^{u}=13.89 \ \mbox{N} \nonumber\\
& \quad \ \ \qquad \underline{\theta}_{i}^{l}=0 \ \mbox{rad}, \ \underline{\theta}_{i}^{u}=4.45 \ \mbox{rad} \nonumber\\
\end{align*}%

Following the procedure described in Sec.~\ref{NumStudWr}, we obtained $\hat{\Sigma}_{\bm{S}}^{c1}=(1e-4)diag([4.771, \ 6.716])$, $\hat{\Sigma}_{\bm{S}}^{c2}=(1e-4)diag([6.019, \ 5.557])$ and $\hat{\Sigma}_{\bm{S}}^{c3}=(1e-4)diag([5.366, \ 5.861])$.

Experimental observations as well as the results obtained using RMT-based model are illustrated in Fig.~\ref{WrenchExprmnt}. Density plots in the second and third columns are obtained by applying kernel smoothing (KS) density estimation technique to the experimental observations (second column) and to the RMT-based wrench samples (third column). Hot colors show higher and cold colors represent lower densities. The results demonstrate the effectiveness of the RMT-based model in capturing the uncertainty. In all three configurations, the mean and variances of the output force obtained using the RMT-based model is adequately close to those obtained from (relatively accurate) experimental measurements.%

\section{Discussion}
\label{Discuss}

We proposed random matrix based models to characterize the uncertainty in robotic platforms. The overal goal of this study was twofold: 1) developing RMT-based models that can provide higher fidelity in capturing (i) motion uncertainty (kinematic analysis) and (ii) wrench uncertainty (static analysis) 2) comprehensive experimental evaluations of the proposed methods to prove the effectiveness and superiority of these models when compared with conventional stochastic models. 

Dealing with motion uncertainty, the Jacobian matrix of the robotic manipulator was modeled as a random matrix. We proposed two RMT-based models based on Wishart perturbation and Gaussian noise matrices. We performed experimental studies using KUKA youBot arm and showed that the RMT-based models outperforms the conventional approaches in capturing the stochastic behavior of the real system. Additionally, in a commonly used particle filtering setup, it was demonstrated that significantly improved estimation of the uncertainty level (error bounds) can be achieved when RMT-based models are used. 

Based on some Gaussianity assumptions, we also proposed an RMT-based model that captures the output wrench uncertainty for which the closed-form expression of the covariance matrix was derived. This was aimed to be an alternative model to a parametric formulation that we developed in our recent work \cite{SoviziTRO}, when detail information on the system parameters is not available in complex multi-agent platforms. We validated the proposed model using an experimental setup consisting of three iRobots that provided a static wrench at the end-effector. It was shown that, if only lower and upper bounds are known (in contrast with the parametric formulation that requires the exact values of the parameters), the RMT-based model can still provide acceptable results in capturing the output wrench uncertainty.   

Following subjects are some of the main extensions to be considered for future extension of the current work. First, RMT-based models can be integrated into motion planning algorithms. Especially when the Gaussian noise matrix model is adopted, a closed-form expression of the motion uncertainty can be augmented to the optimization problem that provides the optimal trajectories. This will result in less uncertain trajectories and hence improved safety and accuracy. The method can also be applied to parallel architecture such as 6-DOF Hexapod (Stewart platform). For the wrench analysis, we provided the required basics, and evaluation was performed on a 2D system. However, applying the method to spatial systems such as cooperative aerial towing platforms is a potential subject that can be considered in future studies.

\bibliographystyle{asmems4}
\bibliography{Sage_LaTeX_Guidelines}

\end{document}